\documentclass[a4paper,fleqn,usenatbib]{mnras}

\usepackage[T1]{fontenc}
\usepackage{ae,aecompl}

\usepackage{graphicx}	
\usepackage{amsmath}	
\usepackage{amssymb}	
\usepackage{footnote}
\usepackage{lscape}
\usepackage{float}            
\usepackage[flushleft]{threeparttable}  

\usepackage{caption}
\usepackage[skip=0.5ex]{subcaption}
\graphicspath{{figures/}}
\usepackage[normalem]{ulem}

\newcommand{\mso}{$\mathcal{M}_{\odot}$}    
\newcommand{\ngc}{$N_\textrm{GC}$}  

\newcommand{\grad}{$^{\circ}$}
\newcommand{\meta}{$[\rm{Fe}/\rm{H}]$}  

\newcommand{\hb}{${\rm H}\beta$}

\usepackage[dvipsnames]{xcolor}

\title[Ages and metallicities of M81 GCs]{Ages and metallicities of globular clusters in M81 using GTC/OSIRIS spectra}

\author[Lomel\'i-N\'u\~nez et al.]{Luis Lomel\'i-N\'u\~nez,$^{1,2}$\thanks{E-mail: luislomeli@ov.ufrj.br (LLN)}
Y.D. Mayya,$^{1}$
L.H. Rodr\'iguez-Merino,$^{1}$
P.A. Ovando,$^{1}$
\newauthor
Jairo A. Alzate,$^{1,3}$
D. Rosa-Gonz\'alez,$^{1}$
B. Cuevas-Otahola,$^{4}$
Gustavo Bruzual,$^{5}$
\newauthor
Arianna Cortesi,$^{2}$
V.M.A
G\'omez-Gonz\'alez$^{6}$
and Carlos G. Escudero$^{7}$
\\
$^{1}$Instituto Nacional de Astrof\'isica \'Optica y Electr\'onica, Luis Enrique Erro 1, Tonantzintla 72840, Puebla, Mexico\\
$^{2}$Valongo Observatory, Federal University of Rio de Janeiro, Ladeira Pedro Antonio 43, Saude Rio de Janeiro, RJ, 20080-090, Brazil\\
$^{3}$Centro de Estudios de F\'isica del Cosmos de Arag\'on\\
$^{4}$Departamento de Matem\'aticas-FCE, Benem\'erita Universidad Aut\'onoma de Puebla\\ 
$^{5}$Instituto de Radioastronom\'ia y Astrof\'isica, UNAM, Campus Morelia, Michoac\'an, M\'exico, C.P. 58089, Mexico\\
$^{6}$Institut f{\"u}r Physik und Astronomie, Universit{\"a}t Potsdam, Karl-Liebknecht-Str. 24/25, 14476 Potsdam, Germany\\
$^{7}$Facultad de Cs. Astron\'micas y Geof\'isicas, UNLP, Paseo del Bosque S/N, B1900FWA, La Plata, Argentina
}

\date{Accepted XXX. Received YYY; in original form ZZZ}

\pubyear{2024}

\begin{document}
\label{firstpage}
\pagerange{\pageref{firstpage}--\pageref{lastpage}}
\maketitle

\begin{abstract}
We here present the results of an analysis of the optical spectroscopy of 42 globular cluster (GC) candidates in the nearby spiral galaxy M81 (3.61~Mpc). The spectra were obtained using the long-slit and MOS modes of the OSIRIS instrument at the 10.4~m Gran Telescopio Canarias (GTC) at a spectral resolution of $\sim$1000. We used the classical H$\beta$ vs [MgFe]$'$ index diagram to separate genuine old GCs from clusters younger than 3 Gyr. Of the 30 spectra with continuum signal-to-noise ratio $>10$, we confirm 17 objects to be classical GCs (age $>10$~Gyr, $-1.4<$[Fe/H]$<-$0.4), with the remaining 13 being intermediate-age clusters (1-7.5~Gyr). We combined age and metallicity data of other nearby spiral galaxies ($\lesssim18$~Mpc) obtained using similar methodology like the one we have used here to understand the origin of GCs in spiral galaxies in the cosmological context. We find that the metal-poor ([Fe/H]<$-$1) GCs continued to form up to 6~Gyr after the first GCs were formed, with all younger systems (age $<8$~Gyr) being metal-rich.

\end{abstract}



\section{Introduction}

Globular Clusters (GCs) are among the oldest objects in the universe, which make them a key component to understand the formation and assembly history of galaxies \citep[]{Brodie:2006}. The availability of wide-field digital detectors on ground-based telescopes and the increased spatial-resolution offered by the Hubble Space Telescope (HST) in the nineties opened up the study of GCs systems in external galaxies \citep[][]{Harris:1991}. Unlike the GCs in the Milky Way, which are selected based on their morphological appearance, the samples of GCs in external galaxies are defined in terms of the photometric properties of compact sources in images \citep[see e.g.][]{Chandar5galaxias:2004, Munoz:2014, Lomeli:2017, Lomeli:2022}. 

\begin{figure} 
    \includegraphics[width=\columnwidth]{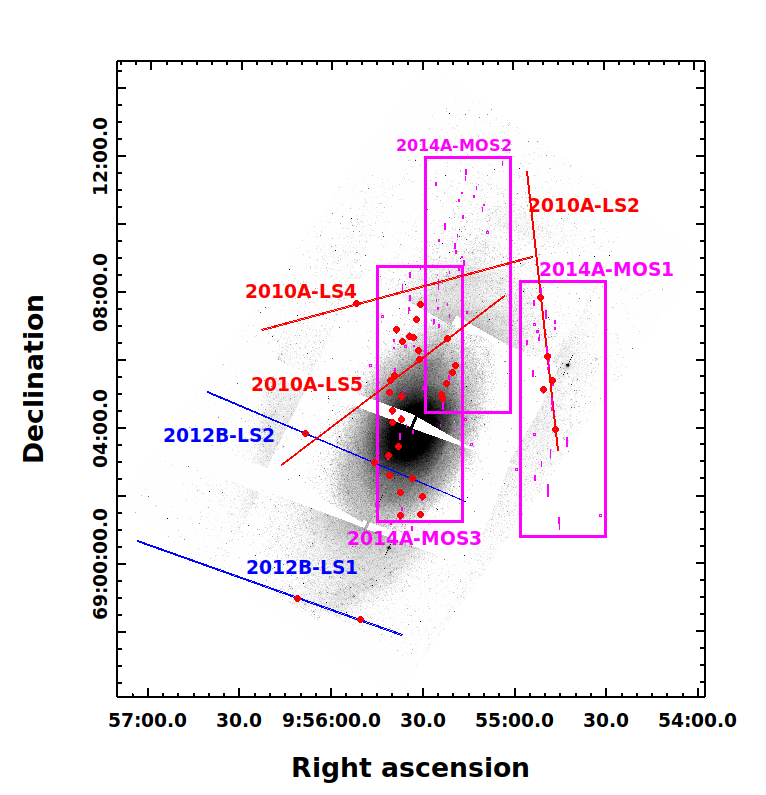}
    \caption{M81 footprint in the HST/ACS $F814W$-band, showing the five long-slits and three MOS fields used for the data analysed in this work. Slits and MOS fields are identified by observing run code listed in Table~\ref{tabla:observaciones}. The GC candidates from S10 are identified by red circles. The slitlets in the MOS fields that do not include GCs are SSCs and suspected Fuzzy Clusters, which are not part of the discussions in this study.}
    \label{figura:footprint} 
\end{figure}

At present, there exists good quality photometric studies of large samples of GCs and spectroscopic studies for a limited number of GCs in elliptical and spiral galaxies.
\citep[see e.g.][]{Brito:2011, Ko:2018, Wang:2021, Kim:2021, Escudero:2022}. 
A common result found in these studies is the bimodal distribution of colours (\citealt{Zepf:1993}, \citealt{Gebhardt:1999}, \citealt{Larsen:2001}). 
A correlation between colour and metallicity was also observed in different galaxies \citep[e.g.][]{Peng:2006, Brito:2011}. Given this correlation, the bimodal colour distribution has been attributed to a bimodality in the abundance of metals \citep[see,][]{Brodie:2006}.
The bimodality would indicate the presence of two different populations of GCs, the metal-poor (commonly referred to as blue), and the metal-rich (or red).
The two kinds of GCs also seem to be spatially segregated, with the distribution of metal-poor blue GCs having a larger scale length as compared to the metal-rich red GCs \citep[e.g.][]{Larsen:2003, Webb:2012, Kartha:2014}.

Two interpretations are in common use to explain the origin of the bimodality of metallicity in GC systems:
a) a major-merger scenario 
proposed by \citet{Ashman:1992}, where the metal-poor GCs are accreted by progenitor merging spiral galaxies and the  metal-rich GCs are formed {\it in-situ} in metal-enriched gas following the merger event,
b) an accretion scenario \citep{Cote:1998} invoking a mass-metallicity relation, 
where the metal-rich GCs were formed {\it in situ} within a massive seed galaxy, while the metal-poor GCs were formed in less-massive satellite galaxies which are subsequently acquired by the host galaxy during the accretion process.
In the major-merger scenario, the metal-poor and metal-rich GCs were formed before and after the merger event, respectively, and hence the metal-poor GCs are expected to be systematically older than the metal-rich GCs. This model predicts an age spread among the GCs. More importantly, clusters of intermediate-age ($\sim$5--10~~Gyr) are expected to be present among the objects classified as GCs from photometric data.
On the other hand, in the accretion scenario both the metal-rich and metal-poor GCs were present in the host galaxies before the merger event. 
Under such circumstances, the relative ages of the metal-rich and metal-poor GCs depend on the formation epochs of low-mass GCs and massive halos \citep[see,][]{Peebles:1984, Rosenblatt:1988}.

Most of the analysis of spectroscopic data of GCs hitherto have been focused on obtaining their metallicity. Modern techniques of analysis of spectral data of unresolved stellar populations allow the determination of ages and metallicities with enough precision to be able to distinguish the predictions of these two cosmological scenarios of galaxy formation. We use new spectroscopic data of the GC systems in M81, a nearby giant spiral (Sab) galaxy located at 3.61~Mpc \citep{Tully:2013}, along with available data on other GC systems, to look for the presence of intermediate-age clusters and also explore the age-metallicity relation in GC systems. 

The population of GCs in M81 has been studied in the past by several authors.
\citet[][]{Perelmuter:1995} found $\sim$70 objects classified as cluster candidates in the inner 11 kpc. In a similar way \citet{Nantaisfoto:2010} and \citet{Mayra:2010} studied the star clusters population in M81 using the available data from the Hubble Space Telescope/Advance Camera for Surveys (HST/ACS), which consisted of 29 adjacent fields covering a field of view of $\sim$340~arcmin$^{2}$ with a sampling of 0.05~arcsec~pixel$^{-1}$ (0.88~pc~pixel$^{-1}$), reporting 233 and 172 GC candidates, respectively. More recently, \citet{Santos:2022} studied a sample of GC candidates around the M81 group (M81, M82 and NGC 3077), using wide-field ground-based images, reporting 642 new GC candidates in a region of 3.5 deg$^{2}$ around the triplet. Some of the GC candidates have been targets of follow-up spectroscopic observations. \citet{Perelmuter_spec:1995} analyzed spectra of 82 GC candidates using the relative strengths of H$\delta$, CaI $\lambda$4227, and FeI $\lambda$4045, confirming 25 of them as bonafide GCs. \citealt{Nantais:2010}, obtained spectra of 74 GCs, and determined metallicities for GCs using an empirical calibration based on Milky Way (MW) GCs. They found a mean metallicity of \meta=$-1.06\pm0.07$~dex.

The main goal of this work is the determination of age and metallicity for the GC candidates in M81 using spectroscopic data with the OSIRIS instrument at the 10.4-m Gran Telescopio Canarias (GTC).
In \S2, we describe the sample of GC candidates, the spectroscopic data, reduction and extraction. We explain the age and metallicity 
determination methods in \S3. In \S4, we carry out a detailed analysis of the determined age and metallicity of the GC candidates. 
In \S5, we combine the M81 data with similar data from the literature for other spiral galaxies to address the age and metallicities 
of the GCs in spiral galaxies in the cosmological context.

\begin{table*}
       \setlength\tabcolsep{9.5pt}
	\centering
    \begin{scriptsize}
	\caption{Log of the long-slit and MOS spectroscopic observations with GTC/OSIRIS in M81.}
	\label{tab:example_table}
	\begin{tabular}{l c c c c c c c c c c} 
        \hline
Run               & PI    &  Date    & PA   & SW  & Exp. time & AM  &  Seeing & Night & Std  & \ngc \\
(1)                    & (2)   &  (3)     & (4)  & (5) & (6)       & (7) &  (8)    &  (9)  & (10) &  (11) \\
         \hline

2010A-LS2  & D. Rosa-Gonz\'alez  & 2010-04-05  & 6.24   & 1.00 & 3$\times$900  & 1.33  & 0.80  & G  & Feige34  & 3  \\
2010A-LS4  & D. Rosa-Gonz\'alez  & 2010-04-05  & 105.20 & 1.00 & 3$\times$900  & 1.56  & 0.80  & G  & Feige34  & 1  \\
2010A-LS5  & D. Rosa-Gonz\'alez  & 2010-04-06  & 127.20 & 1.00 & 3$\times$900  & 1.43  & 0.80  & G  & Feige34  & 1  \\
2012B-LS1  & Y. D. Mayya      &  2013-01-12    & 250.50 & 1.23 & 3$\times$1500 & 1.31  & 0.79  & D  & Feige34  & 2  \\
2012B-LS2  & Y. D. Mayya      &  2013-01-12    & 247.00 & 1.23 & 3$\times$1500 & 1.40  & 0.97  & D  & Feige34  & 2  \\
2014A-MOS1 & Y. D. Mayya      &  2014-04-03    & 0.00   & 1.20 & 3$\times$1308 & 1.31  & 0.90  & D  & Ross 640 & 5  \\
2014A-MOS2 & Y. D. Mayya      &  2014-03-23    & 0.00   & 1.20 & 3$\times$1308 & 1.35  & 1.00  & D  & Ross 640 & 8  \\
2014A-MOS3 & Y. D. Mayya      &  2014-04-03    & 0.00   & 1.20 & 3$\times$1308 & 1.34  & 0.80  & D  & Ross 640 & 25  \\
	\hline
	\end{tabular}
            \begin{tablenotes}
                \begin{small}
                \item {\it Notes:} (1) Observation run. (2) Principal investigator (PI). (3) Observational date (year-month-day). (4) Position angle (\grad) of the slit as measured on the astrometrized image. (5) Slit-width ($\arcsec$). (6) Exposure time (number of exposures $\times$ integration time in seconds). (7) Mean airmass of the three integrations. (8) Seeing ($\arcsec$). (9) Night (G$=$grey or D$=$dark), clear skies (cirrus reported only for 2010A-LS5). (10) Standard star name. (11) GCs number in each observation campaign.
                \end{small}
           \end{tablenotes}
        \label{tabla:observaciones}
    \end{scriptsize}
\end{table*}

\section{Sample and spectroscopic data}

\citet{Mayra:2010} (hereafter S10) analyzed HST images of M81 using multi pointing observations
with the Advanced Camera for Surveys (ACS) in $F435W$ ($B$), $F606W$ ($V$) and $F814W$ ($I$) bands.
The population of stellar clusters was classified in two subpopulations: 
young (blue) 
3~Myr to 2~Gyr and old (red) $\gtrsim$5~Gyr. 
They obtained 263 blue and 172 red star clusters. 
The selection criteria of stellar clusters was based on structural parameters obtained with SExtractor\footnote{\url{https://www.astromatic.net/software/sextractor}} \citep{Sextractor:1996}: {\sc fwhm}, {\sc area} and {\sc ellipticity}; and an evolutionary parameter, the photometric colour.

We carried out a spectroscopic campaign to observe a subsample of star clusters catalogued by S10 using the OSIRIS instrument at the 10.4-m Gran Telescopio Canarias (GTC). The observational campaign consisted of three runs, one each in 2010, 2012 and 2014, totalling
eight independent pointings. The footprints of these pointings are shown in Figure~\ref{figura:footprint}, with the observational details for each pointing given in Table~\ref{tabla:observaciones}. 
The 2010 and 2012 observations were carried out using long-slits whereas the 2014 observations were carried out in the MOS mode.
The number of GCs \ngc\ in each pointing is given in column 11 in Table~\ref{tabla:observaciones}.
The targets for observations included blue objects, which were super star cluster (SSCs), GCs and suspected Fuzzy clusters. 
The pointing 2014A-MOS3 was specially designed for the observation of GC candidates, where 25 GCs were observed. 
The pointing 2014A-MOS1 and 2014A-MOS3 included blue objects, which were SSCs and suspected Fuzzy clusters.
In these different observational runs, 47 GC candidates were targeted. However, five of them were too close to the slit borders for a reliable spectral extraction. Hence, the final sample of GC candidates with extracted spectra contained 42 objects.
Results obtained from spectra targeted towards SSCs helped to detect individal Wolf Rayet stars in M81 (see \citealt{Mauricio:2016}).
Some slits passed through nebular regions which helped to obtain nebular abundance and its gradient in M81 (see \citealt{Arellano:2016}). Spectra of the brightest GC, M81-GC1 was analyzed by \citet{Mayya:2013}.

\subsection{Spectroscopic observational details}

Our sample has spectra of 42 of the 172 GC candidates reported in S10. All observations were carried out with the OSIRIS spectrograph at the Nasmyth-B focus using the R1000B grism with a slit-width 1.0 to 1.23~arcsec, covering a spectral range from 3700 to 7500~\AA\ with a spectral resolution of $\sim$7~\AA\ measured at $\lambda_{\rm{C}}=5455$~\AA. We selected the observation mode with a binning of 2$\times$2, obtaining a spatial scale of 0.254~arcsec~pixel$^{-1}$ (in the horizontal axis), and a spectral sampling of $\sim$2~\AA~pixel$^{-1}$ (in the vertical axis). In Table 1, we summarize the observational specifications: Observational run (column 1), PI (column 2), observation date (column 3), slit position angle (column 4), slit-width in arcsec (column 5), exposure time (column 6), air mass (column 7), observational seeing (column 8), sky type (column 9), standard stars (column 10) and \ngc\ in each campaign observation (column 11).

In Table \ref{tabla:muestra}, we tabulate the photometric and spectroscopic parameters, 
of our sample of GC candidates. Object IDs (column 1) are taken from \citet{Lomeli:2022} (hereafter L22).
Coordinates (RA, DEC) in J2000
epoch (column 2-3), $B$-magnitude (column 4) and $(B-I)_{0}$ colour (column 5) taken from S10, $(u-g)_{0}$ colour (column 6) are taken from L22, observed velocity (column 7), root-mean-square (RMS) and signal-to-noise ratio (SNR) of the continuum are measured in our spectra at 4190~\AA\ (column 8-9). Quality (Q) of spectra of each GC candidate, which is obtained by visual inspection and is mainly based on a combination of the SNR, sky subtraction accuracy and position of the object in the slitlet (10). A more detailed description about Q parameter is given in Section~\ref{section:spectral_quality}. Observation run for each GC candidate (11), and its
name from S10 (column 12) are also given.

\begin{table*}  
\setlength\tabcolsep{2.5pt} 
\centering   
\caption{Sample of GC candidates.} 
\begin{scriptsize}
\label{tabla:muestra}   
\begin{tabular}{l l l c c c r r r r l r}  
\hline    
Name$^\ddagger$ & RA(J2000)$^\dagger$ & DEC(J2000)$^\dagger$ & $F435W_{0}$$^\dagger$ & $(F435W-F814W)_{0}$$^\dagger$ & $(u-g)_{0}^\ddagger$& v$_{obs}$ & RMS & SNR & Quality   & Run  & Object$^\dagger$ \\  
  & [h:m:s]       & [d:m:s]        &  [mag]        & [mag]  &  [mag] & [km~s$^{-1}$] & [erg cm$^{-2}$ s$^{-1}$ \AA]       &   &    &  &  \\
(1)  & (2)       & (3)        & (4)         & (5)                 &  (6) & (7) & (8)       & (9)  & (10)   & (11)  & (12)  \\  
\hline 
GC1   &  9:55:21.880   &  +69:06:37.88 &  17.83  &  2.06  &  1.63$\pm$0.10   &  152.60 &  4.33e-17 &  22.10  &   G  &  2010A-LS5  &   R05R06584   \\ 
GC3   &  9:55:39.975   &  +69:04:10.37 &  18.59  &  1.81  &  1.43$\pm$0.10   &  372.30 &  1.41e-17 &  34.80  &   G  &  2014A-MOS3  &   R10R03509   \\ 
GC4   &  9:56:08.387   &  +69:03:51.35 &  18.76  &  2.16  &  1.82$\pm$0.10   &  -218.97 &  4.51e-18 &  25.70  &   G  &  2012B-LS2  &   R10R10692   \\ 
GC5   &  9:55:45.870   &  +69:03:00.70 &  18.77  &  2.24  &  1.47$\pm$0.10   &  -192.60 &  6.04e-18 &  30.90  &   G  &  2012B-LS2  &   R10R09559   \\ 
GC7   &  9:55:22.028   &  +69:05:19.13 &  19.14  &  2.12  &  1.71$\pm$0.10   &  -20.57 &  9.42e-18 &  27.30  &   G  &  2014A-MOS1  &   R05R10583   \\ 
GC8   &  9:56:11.101   &  +68:59:00.67 &  19.38  &  2.28  &  1.68$\pm$0.10   &  -18.51 &  1.09e-18 &  6.80  &   B  &  2012B-LS1  &   R13R19709   \\ 
GC9   &  9:55:37.245   &  +69:02:07.77 &  19.51  &  2.34  &  2.18$\pm$0.10   &  -90.06 &  3.75e-18 &  34.00  &   G  &  2014A-MOS3  &   R09R02463   \\ 
GC10  &  9:55:54.472   &  +69:02:52.80 &  19.56  &  2.21  &  1.55$\pm$0.10   &  -10.51 &  3.19e-18 &  36.90  &   G  &  2014A-MOS3  &   R10R11944   \\ 
GC11  &  9:55:37.763   &  +69:03:28.16 &  19.56  &  1.80  &  1.85$\pm$0.10   &  152.34 &  2.88e-18 &  40.70  &   G  &  2014A-MOS3  &   R10R05960   \\ 
GC13  &  9:55:32.809   &  +69:06:40.01 &  19.72  &  2.07  &  1.67$\pm$0.10   &  -127.04 &  3.43e-18 &  27.60  &  {\bf G}   &  2014A-MOS3  &   R06R15013   \\ 
GC16  &  9:55:51.816   &  +69:07:39.74 &  19.88  &  2.27  &  1.64$\pm$0.10   &  -82.26 &  5.38e-18 &  23.60  &   G  &  2010A-LS4  &   R06R03030   \\ 
GC27  &  9:55:57.695   &  +69:02:23.34 &  20.41  &  2.27  &  1.54$\pm$0.10   &  -107.85 &  2.45e-18 &  22.20  &   G  &  2014A-MOS3  &   R10R14199   \\ 
GC29  &  9:55:50.181   &  +68:58:22.91 &  20.57  &  2.02  &  1.44$\pm$0.10   &  -220.98 &  1.36e-18 &  30.00  &   {\bf G}   &  2012B-LS1  &   R12R17286   \\   
GC32  &  9:55:39.307   &  +69:05:32.87 &  20.64  &  1.81  &  2.08$\pm$0.10   &  149.25 &  3.95e-18 &  14.00  &   B  &  2014A-MOS3  &   R06R14067   \\ 
GC44  &  9:55:19.182   &  +69:05:50.35 &  20.98  &  2.50  &  1.92$\pm$0.10   &  279.33 &  1.94e-18 &  12.80  &   G  &  2014A-MOS1  &   R05R08459   \\ 
GC46  &  9:55:31.034   &  +69:06:02.44 &  21.04  &  2.44  &  1.70$\pm$0.10   &  209.01 &  1.84e-18 &  13.50  &   G  &  2014A-MOS1  &   R06R17963   \\ 
GC47  &  9:55:40.447   &  +69:05:24.98 &  21.06  &  2.60  &  3.03$\pm$0.10   &  201.64 &  1.25e-18 &  24.10  &   {\bf G}   &  2014A-MOS3  &   R06R13765   \\ 
GC55  &  9:54:46.269   &  +69:03:57.49 &  21.16  &  2.29  &  2.09$\pm$0.10   &  30.60 &  2.54e-18 &  6.00  &   B  &  2010A-LS2  &   R04R10733   \\ 
GC59  &  9:55:38.468   &  +69:06:55.29 &  21.28  &  2.96  &  2.36$\pm$0.10   &  408.48 &  1.39e-18 &  16.50  &   G  &  2014A-MOS3  &   R06R11333   \\ 
GC60  &  9:55:36.700   &  +69:06:33.18 &  21.29  &  2.42  &  2.00$\pm$0.10   &  278.55 &  9.79e-19 &  21.10  &   G  &  2014A-MOS3  &   R06R13269   \\ 
GC61  &  9:55:41.294   &  +69:03:11.48 &  21.35  &  2.28  &  1.54$\pm$0.10   &  -103.77 &  2.15e-18 &  12.10  &   G  &  2014A-MOS3  &   R10R07655   \\ 
GC62  &  9:55:20.134   &  +69:05:37.88 &  21.37  &  2.12  &  1.27$\pm$0.10   &  184.19 &  1.14e-18 &  12.60  &   {\bf G}   &  2014A-MOS1  &   R05R09834   \\ 
GC66  &  9:54:51.016   &  +69:07:50.33 &  21.44  &  2.66  &  1.78$\pm$0.10   &  203.25 &  1.18e-18 &  21.10  &   G  &  2014A-MOS1  &   R02R14402   \\ 
GC70  &  9:55:23.753   &  +69:04:59.70 &  21.54  &  1.96  &  2.43$\pm$0.10   &  121.53 &  4.53e-18 &  27.80  &   G  &  2014A-MOS1  &   R05R11972   \\ 
GC79  &  9:55:30.063   &  +69:01:59.74 &  21.75  &  2.19  &  1.87$\pm$0.10   &  86.49 &  1.39e-18 &  11.20  &   G  &  2014A-MOS3  &   R09R02775   \\ 
GC80  &  9:55:34.372   &  +69:06:42.49 &  21.75  &  1.89  &  2.97$\pm$0.10   &  129.54 &  1.38e-18 &  28.40  &   G  &  2014A-MOS1  &   R06R14078   \\ 
GC82  &  9:55:40.744   &  +69:02:37.68 &  21.75  &  2.05  &  1.74$\pm$0.10   &  -225.51 &  1.45e-18 &  10.10  &   B  &  2014A-MOS3  &   R09R03511   \\ 
GC87  &  9:54:50.267   &  +69:05:08.63 &  21.89  &  2.03  &  2.36$\pm$0.10   &  -3.79 &  5.99e-19 &  19.30  &   {\bf G}   &  2014A-MOS1  &   R04R02765   \\ 
GC91  &  9:55:36.792   &  +69:04:55.56 &  21.99  &  1.90  &  1.45$\pm$0.10   &  -95.40 &  1.12e-18 &  9.10  &   B  &  2014A-MOS3  &   R06R15901   \\ 
GC98  &  9:55:30.647   &  +69:01:28.96 &  22.18  &  2.14  &  1.77$\pm$0.10   &  -214.17 &  9.48e-19 &  4.00  &   B  &  2014A-MOS3  &   R09R01598   \\ 
GC103 &  9:55:40.954   &  +69:05:04.11 &  22.25  &  2.14  &  1.97$\pm$0.10   &  94.11 &  2.15e-18 &  30.30  &   {\bf G}   &  2014A-MOS3  &   R06R14042   \\ 
GC104 &  9:55:23.335   &  +69:04:52.76 &  22.25  &  2.36  &  1.02$\pm$0.10   &  145.74 &  2.04e-18 &  30.20  &   G  &  2014A-MOS2  &   R05R12224   \\ 
GC111 &  9:55:33.429   &  +69:02:32.06 &  22.70  &  2.31  &  1.23$\pm$0.10   &  24.68 &  9.90e-19 &  6.10  &   B  &  2014A-MOS3  &   R09R03952   \\ 
GC114 &  9:55:31.260   &  +69:06:18.35 &  22.83  &  2.45  &  1.36$\pm$0.10   &  174.69 &  1.51e-18 &  27.60  &   G  &  2014A-MOS1  &   R06R15830   \\ 
GC123 &  9:55:37.304   &  +69:01:25.73 &  23.11  &  1.95  &  0.78$\pm$0.10   &  -68.12 &  3.65e-18 &  7.70  &   B  &  2014A-MOS3  &   R09R01067   \\ 
GC125 &  9:55:31.847   &  +69:07:12.65 &  23.17  &  2.80  &  1.57$\pm$0.10   &  133.18 &  7.25e-19 &  6.90  &   B  &  2014A-MOS3  &   R06R14182   \\ 
GC134 &  9:55:50.260   &  +69:02:42.93 &  23.56  &  2.65  &  0.39$\pm$0.10   &  -83.58 &  1.84e-18 &  2.20  &   B  &  2014A-MOS3  &   R10R11530   \\ 
GC135 &  9:55:30.759   &  +69:07:38.95 &  19.05  &  2.05  &  0.49$\pm$0.10   &  100.05 &  2.56e-18 &  50.60  &   G  &  2014A-MOS3  &   R06R14272   \\ 
GC136 &  9:54:47.159   &  +69:05:24.36 &  20.66  &  2.04  &  1.43$\pm$0.10   &  -88.89 &  2.88e-18 &  15.50  &   {\bf G}   &  2014A-MOS1  &   R05R04684   \\ 
GC146 & 9:54:48.938    & +69:06:05.74  &  22.53  &  1.95  &  -   &  114.87 &  1.84e-18 &  2.00  &   B  &  2010A-LS2  &   R05R01311   \\ 
GC159 &  9:55:37.041   &  +69:04:15.70 &  20.86  &  -  &  2.30$\pm$0.10   &  193.66 &  9.62e-18 &  33.80  &   G  &  2014A-MOS3  &   R10R02298   \\ 
GC160 &  9:55:39.966   &  +69:04:32.48 &  21.00  &  -  &  1.70$\pm$0.10   &  -90.48 &  1.47e-18 &  13.40  &   B  &  2014A-MOS3  &   R10R01848   \\ 
\hline   
\end{tabular}   
\begin{tablenotes}   
\begin{scriptsize} 
\item { 
(1) GC candidate name. 
(2,3) Right ascension and declination coordinates (J2000). 
(4) $B$-magnitude from HST-$F435W$ band corrected for galactic extinction.
(5) Colour $(B-I)_{0}$, corrected for galactic extinction.
(6) Colour $(u-g)$, corrected for galactic extinction.
(7) Observed mean velocity estimated with H$\alpha$ and H$\beta$ absorption lines, v$_{obs}$ in km~s$^{-1}$. }
(8,9) RMS and SNR/pixel measured at 4190~\AA\ in the continuum. 
(10) Quality (Q) parameter of spectra was assigned by visual inspection: B=bad, G=good.
(11) Observation run for each GC candidate.
(12) Name of GC candidates from S10.
\item {$^\dagger$ Values from S10.}
\item {$^\ddagger$ Values from L22.}
\end{scriptsize}   
\end{tablenotes} 
\end{scriptsize}
\end{table*}

\subsection{Data reduction and spectral extraction}

Both the long-slit and MOS spectral images were reduced using GTCMOS\footnote{\url{https://www.inaoep.mx/~ydm/gtcmos/gtcmos.html}}, which is a dedicated pipeline devoted to the reduction of OSIRIS/GTC observational blocks (OBs). The main steps in the reduction process are the following:
(i) joining the two CCD images into a single mosaic image; ii) bias subtraction; iii) wavelength calibration 
and iv) flux calibration using standard stars (column (10) of Table \ref{tabla:observaciones}).
Wavelength calibration was carried out for each slitlet using the arc lamp images using the same slit mask as that used for source observations.
The final product is a 2D-spectra corrected for geometric distortions, wavelength and flux calibrated. For a more detailed description of the software and reduction method see \citet{Mauricio:2016}.

\begin{figure}
    \includegraphics[height=5cm,width=8.5cm]{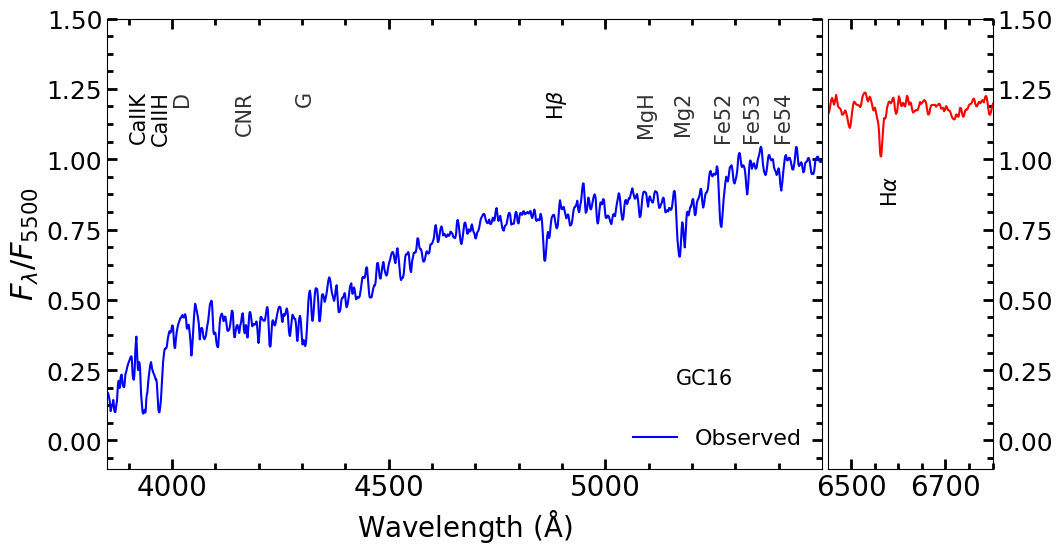} 
    \caption{Example of the extracted spectrum of a GC candidate. In this case we show the GC named GC16. 
    The observed flux (normalized at 5500~\AA) from: i) 3900 to 6000~{\AA} ({\it blue line}) and ii) 6450 to 6800~{\AA} ({\it red line}) is plotted.
    }
    \label{figura:espectro_observado}
\end{figure}

Given the relative faintness of the extragalactic GCs compared to the combined brightness of sky and disk background, it is crucial for the purpose of this study to maximise the SNR ratio while extracting 1-D spectra from the wavelength and flux calibrated 2D-images.
Hence each of the 42 spectra was extracted interactively using the IRAF task $apall$. In all cases, the continuum was strong enough for a reliable tracing of the spectrum. Extraction of the spectra was carried out using a fixed window size of $\sim$4 pixels (1 arcsec) centered on the trace axis. For each object, a sky+background spectrum is extracted in windows of 3 to 6 pixel widths on either side of the object, which is then scaled to the size of object extraction window and subtracted to get pure GC spectra. For MOS observations, the background window was chosen in slitlets specially reserved for that purpose, when the slitlet containing the object did not have enough background pixels. In either case, we ensured by examining visually the HST/ACS multiband images that the surface brightness of the disk in the zone selected for the extraction of the sky+background spectrum is as close to that in the vicinity of the target GC.

In Figure \ref{figura:espectro_observado} we show an example of an extracted spectrum of a GC candidate, where we indicate the position of some of the prominent features observed in GC spectra.
For the sake of clarity, the spectrum is shown in two sections: the bluer range (3900-6000~\AA) which has most of the GC features (CaIIK, CaIIH, CNR, G-band, MgH, Mg2, Mgb, Fe52, Fe53, and F54), and a red window (6450-6800~\AA) for the H$\alpha$ line. The undisplayed middle part is featureless in all cases. The H${\alpha}$ and H${\beta}$ are used to obtain an average radial velocity of each GC candidate. All spectra shown in this work are in the rest frame after correcting for the Doppler shift using the average radial velocity.
\begin{figure}
\begin{center}
    \includegraphics[width=0.8\columnwidth]{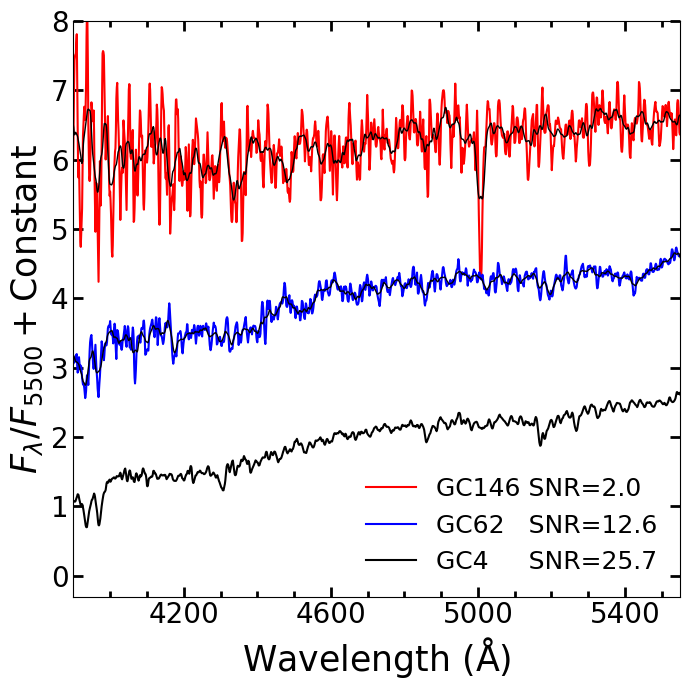}
    \caption{Comparison of spectral quality:
    spectrum with low SNR (red line), 
    medium SNR (blue line) and
    high SNR (black line).
    In the cases of low and medium SNR we superposed their smoothed spectra in black color.  }
    \label{figura:calidad} 
\end{center}
\end{figure}

\subsection{{\bf Spectral quality}}
\label{section:spectral_quality}

In the whole sample we have 9 objects with SNR at 4190~\AA$<$10, 11 objects with 10$\leq$SNR$\leq$20, and 22 with SNR$>$20. 
In Figure~\ref{figura:calidad}, we show examples for three values (low, intermediate and high) of continuum SNR. The spectra were normalized to their flux at 5500~\AA\ and vertically shifted adding an arbitrary constant. For the medium and low SNR spectra, we also show smoothed versions of the
spectra. It can be noted that the absorption features can be identified in the smoothed spectrum even though they are buried in the
noise in the case of low SNR spectrum. On the other hand, absorption features are identifiable without the need for smoothing for SNR>10.
We assigned a quality index of Good (G) or Bad (B) for each extracted spectrum, which is based on the SNR, accuracy of sky subtraction and the
positioning of the object in the slitlet. For a spectrum to be assigned a G quality, it has to have an SNR of at least 10. All spectra with SNR<10,
and some spectra with SNR>10, are assigned a B quality if the sky subtracted spectrum showed unphysical continuum shape (eg., wavy nature, negative fluxes etc.), or the object was at the 
edge of a slitlet, in which case the spectrum does not cover the entire targetted wavelength range. In column (10) of Table~\ref{tabla:muestra} we list the spectral quality for each GC spectrum. In total, spectra for 12 GCs with Quality B are not useful for our analysis. Thus the spectroscopic sample consists of 30 objects with Quality parameter G.

\begin{figure}
    \includegraphics[width=0.9\columnwidth]{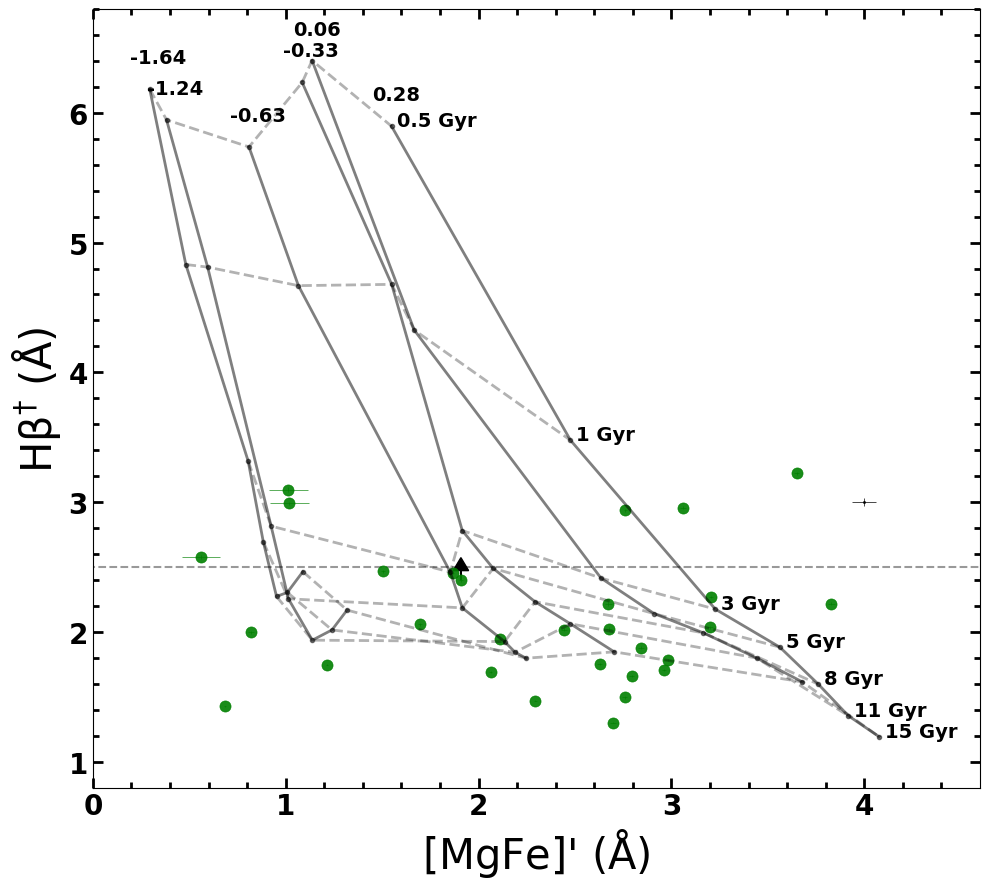}
    \caption{Index-index diagram H$\beta$$^{\dagger}$ vs [MgFe]$'$ for the sample of GC candidates (green solid dots). 
    The grid represent the SSP models from \citet{Bruzual:2003}, with metallicities Z=0.0004, 0.001, 0.004, 0.008, 0.019, and 0.05 and ages 0.5, 1, 3, 5, 8, 11 and 15 Gyr.  
    {\it Black dashed line} is the limit of $\sim$3~Gyr to separate GCs from young intermediate reddened cluster.}
    \label{figura:hbeta_mgfe} 
\end{figure}

\begin{table}
       \setlength\tabcolsep{3.0pt}
    \begin{small}
	\centering
	\caption{Spectral indices. }
	\label{tabla:indices_lino}
	\begin{tabular}{c c c c c c c c c c} 
        \hline
        Index    & BC         & I        & RC         \\ 
        name     & [\AA]      & [\AA]    & [\AA]            \\ 
        (1)      & (2)        & (3)    & (4)  \\
        \hline
        CNR  & 4082.00 4118.50 & 4144.00 4177.50 & 4246.00 4284.75  \\
        G    & 4268.25 4283.25 & 4283.25 4317.00 & 4320.75 4335.75  \\
        MgH  & 4897.00 4958.00 & 5071.00 5134.75 & 5303.00 6366.75  \\
        Mg2  & 4897.00 4958.00 & 5156.00 5197.25 & 5303.00 6366.75  \\
        Mgb  & 5144.50 5162.00 & 5162.00 5193.25 & 5193.25 5207.00  \\
        Fe52 & 5235.50 5249.25 & 5248.00 5286.75 & 5288.00 5319.25  \\
        Fe53 & 5307.25 5317.25 & 5314.75 5353.50 & 5356.00 5364.75  \\
        Fe54$^\dagger$ & 5376.25 5387.50 & 5387.50 5415.00 & 5415.00 5425.00  \\
	\hline
	H$\beta^\dagger$ & 4827.875 4847.875 & 4847.875 4876.625 & 4876.625 4891.625 \\
	Mgb$^\dagger$  & 5142.625 5161.375 & 5160.125 5192.625 & 5191.375 5206.375 \\
	Fe52$^\dagger$ & 5233.150 5248.150 & 5245.650 5285.650 & 5285.650 5318.150 \\
	Fe53$^\dagger$ & 5304.625 5315.875 & 5312.125 5352.125 & 5353.375, 5363.375    \\
	\hline
	\end{tabular}
           \begin{tablenotes}
                \begin{scriptsize}
                \item {\it Notes:} 
                $\dagger$ Index definition from \citet{Trager:1998}.
                \end{scriptsize}
           \end{tablenotes}
        \end{small}
\end{table}

\section{Analysis}
\label{seccion:analisis}

In this section, we describe in detail the methods we have used to determine the metallicity and ages of the GC candidates.

\begin{center}            
\begin{table}
    \setlength\tabcolsep{3.0pt}
    \begin{scriptsize} 
	\caption{Spectral indices from \citet{Trager:1998} and [MgFe]$'$. }
	\label{tabla:ancho_equivalente}
	\begin{tabular}{l r r r r r}
\hline  
GC  &  H$\beta$$^{\dagger}$  &   Mgb$^{\dagger}$  &   Fe52$^{\dagger}$  &  Fe53$^{\dagger}$  & [MgFe]$'$ \\ 
    &  [\AA]     &  [\AA] &  [\AA]  &  [\AA] & [\AA]  \\       
(1) &  (2)       &   (3)  &   (4)  &  (5)    & (6)  \\ 
\hline
GC1  &    2.02$\pm$0.010   &    2.81$\pm$0.012  &    2.19$\pm$0.010  &     1.97$\pm$0.005  &     2.44$\pm$0.014  \\  
GC3  &    1.75$\pm$0.006   &    1.32$\pm$0.003  &    1.24$\pm$0.001  &     0.80$\pm$0.002  &     1.22$\pm$0.005  \\  
GC4  &    1.66$\pm$0.008   &    3.41$\pm$0.009  &    2.42$\pm$0.010  &     1.96$\pm$0.005  &     2.79$\pm$0.011  \\  
GC5  &    1.78$\pm$0.009   &    3.55$\pm$0.012  &    2.68$\pm$0.010  &     2.06$\pm$0.006  &     2.98$\pm$0.015  \\  
GC7  &    1.75$\pm$0.009   &    3.04$\pm$0.010  &    2.29$\pm$0.010  &     2.22$\pm$0.005  &     2.63$\pm$0.012  \\  
GC9  &    1.88$\pm$0.011   &    3.51$\pm$0.014  &    2.29$\pm$0.010  &     2.33$\pm$0.006  &     2.84$\pm$0.016  \\  
GC10  &    1.95$\pm$0.011   &    2.45$\pm$0.009  &    1.82$\pm$0.001  &     1.80$\pm$0.004  &     2.11$\pm$0.011  \\  
GC11  &    2.99$\pm$0.010   &    1.05$\pm$0.004  &    0.81$\pm$0.001  &     1.45$\pm$0.002  &     1.02$\pm$0.006  \\  
GC13  &    2.06$\pm$0.010   &    1.90$\pm$0.008  &    1.48$\pm$0.001  &     1.57$\pm$0.003  &     1.70$\pm$0.009  \\  
GC16  &    2.02$\pm$0.012   &    3.51$\pm$0.012  &    2.05$\pm$0.010  &     2.02$\pm$0.005  &     2.68$\pm$0.016  \\  
GC27  &    1.50$\pm$0.010   &    3.68$\pm$0.009  &    2.17$\pm$0.010  &     1.79$\pm$0.004  &     2.76$\pm$0.013  \\  
GC29  &    1.43$\pm$0.004   &    0.62$\pm$0.006  &    0.67$\pm$0.001  &     0.96$\pm$0.003  &     0.69$\pm$0.007  \\  
GC44  &    3.22$\pm$0.009   &    4.12$\pm$0.009  &    2.99$\pm$0.010  &     3.84$\pm$0.007  &     3.65$\pm$0.014  \\  
GC46  &    2.22$\pm$0.011   &    4.80$\pm$0.014  &    2.95$\pm$0.010  &     3.32$\pm$0.005  &     3.83$\pm$0.017  \\  
GC47  &    2.58$\pm$0.013   &    0.53$\pm$0.007  &    0.71$\pm$0.001  &     0.26$\pm$0.001  &     0.56$\pm$0.008  \\  
GC59  &    2.46$\pm$0.010   &    2.35$\pm$0.006  &    1.74$\pm$0.001  &     0.80$\pm$0.003  &     1.86$\pm$0.008  \\  
GC60  &    2.96$\pm$0.012   &    3.78$\pm$0.010  &    2.61$\pm$0.010  &     2.13$\pm$0.005  &     3.06$\pm$0.014  \\  
GC61  &    1.30$\pm$0.006   &    3.17$\pm$0.009  &    2.29$\pm$0.010  &     2.31$\pm$0.006  &     2.70$\pm$0.014  \\  
GC62  &    2.00$\pm$0.012   &    0.34$\pm$0.012  &    2.28$\pm$0.010  &     1.18$\pm$0.008  &     0.82$\pm$0.016  \\  
GC66  &    2.03$\pm$0.012   &    3.92$\pm$0.013  &    2.54$\pm$0.010  &     2.77$\pm$0.006  &     3.20$\pm$0.016  \\  
GC70  &    1.70$\pm$0.007   &    3.52$\pm$0.009  &    2.49$\pm$0.010  &     2.50$\pm$0.005  &     2.96$\pm$0.011  \\  
GC79  &    2.22$\pm$0.015   &    2.87$\pm$0.011  &    2.83$\pm$0.010  &     1.59$\pm$0.004  &     2.67$\pm$0.017  \\  
GC80  &    1.69$\pm$0.008   &    2.36$\pm$0.004  &    1.89$\pm$0.010  &     1.56$\pm$0.003  &     2.06$\pm$0.007  \\  
GC87  &    3.10$\pm$0.019   &    0.63$\pm$0.005  &    2.10$\pm$0.020  &     0.35$\pm$0.006  &     1.01$\pm$0.019  \\  
GC103  &        -            &    2.10$\pm$0.011  &    1.98$\pm$0.001  &     1.63$\pm$0.004  &     1.99$\pm$0.012  \\  
GC104  &    2.27$\pm$0.009   &    3.68$\pm$0.007  &    2.89$\pm$0.010  &     2.51$\pm$0.005  &     3.21$\pm$0.010  \\  
GC114  &    2.94$\pm$0.011   &    3.23$\pm$0.005  &    2.36$\pm$0.010  &     2.35$\pm$0.004  &     2.76$\pm$0.009  \\  
GC135  &    2.47$\pm$0.012   &    1.52$\pm$0.007  &    1.50$\pm$0.001  &     1.46$\pm$0.004  &     1.50$\pm$0.009  \\  
GC136  &    2.40$\pm$0.016   &    -               &    3.74$\pm$0.010  &     3.37$\pm$0.004  &     1.91$\pm$0.011  \\  
GC159  &    1.47$\pm$0.004   &    2.75$\pm$0.005  &    2.02$\pm$0.010  &     1.62$\pm$0.003  &     2.29$\pm$0.008  \\   
\hline
        \end{tabular}
           \begin{tablenotes}
                \begin{small}
                \item {\it Notes:} 
                (1) GC name. (2) H${\beta}^{\dagger}$. (3) Mgb$^{\dagger}$. (4) Fe52$^{\dagger}$. (5) Fe53$^{\dagger}$.
                \end{small}
           \end{tablenotes}
        \end{scriptsize}
\end{table}
\end{center}

\begin{center}            
\begin{center}            
\begin{table*}
    \setlength\tabcolsep{4.8pt}
        \begin{scriptsize} 
	\caption{Metallicities and spectral indices measured. }
	\label{tabla:indices_medidos}
	\begin{tabular}{l r r r r r r r r r r r}
\hline
Name &        [Fe/H]&   CNR&     G& MgH&   Mg2&   Mgb&   Fe52&   Fe53&   Fe54 \\
     &        [dex] & [mag]& [mag]& [mag]& [mag]&   [mag]&   [mag]&   [mag] & [mag] \\

(1)&        (2)&     (3)&   (4)&      (5)&   (6)&   (7)&   (8)&   (9)&   (10) \\
\hline
GC1    &  -0.584$\pm$0.319   &  0.047$\pm$0.009  &  0.129$\pm$0.027  &  0.061$\pm$0.008  &   0.157$\pm$0.015  &   0.105$\pm$0.019  &   0.062$\pm$0.012  &  0.052$\pm$0.008  &  0.055$\pm$0.004& \\  
GC3    &  -1.374$\pm$0.254   &  -0.024$\pm$0.004  &  0.083$\pm$0.008  &  0.064$\pm$0.002  &   0.118$\pm$0.006  &   0.040$\pm$0.004  &   0.035$\pm$0.004  &  0.013$\pm$0.005  &  0.024$\pm$0.003& \\  
GC4    &  -0.613$\pm$0.275   &  0.044$\pm$0.010  &  0.191$\pm$0.030  &  0.055$\pm$0.007  &   0.158$\pm$0.015  &   0.132$\pm$0.015  &   0.069$\pm$0.010  &  0.052$\pm$0.008  &  0.052$\pm$0.004& \\  
GC5    &  -0.416$\pm$0.305   &  0.090$\pm$0.008  &  0.186$\pm$0.033  &  0.064$\pm$0.007  &   0.175$\pm$0.015  &   0.132$\pm$0.017  &   0.078$\pm$0.012  &  0.055$\pm$0.009  &  0.063$\pm$0.004& \\  
GC7    &  -0.515$\pm$0.258   &  0.067$\pm$0.008  &  0.164$\pm$0.025  &  0.094$\pm$0.007  &   0.174$\pm$0.010  &   0.114$\pm$0.015  &   0.068$\pm$0.010  &  0.064$\pm$0.006  &  0.055$\pm$0.004& \\ 
GC9    &  -0.182$\pm$0.334   &  0.076$\pm$0.012  &  0.155$\pm$0.033  &  0.069$\pm$0.009  &   0.211$\pm$0.014  &   0.138$\pm$0.023  &   0.075$\pm$0.014  &  0.069$\pm$0.009  &  0.080$\pm$0.006& \\
GC10   &  -0.976$\pm$0.256   &  0.008$\pm$0.013  &  0.159$\pm$0.021  &  0.032$\pm$0.007  &   0.134$\pm$0.010  &   0.092$\pm$0.015  &   0.052$\pm$0.010  &  0.051$\pm$0.005  &  0.033$\pm$0.003& \\  
GC11$^\dagger$   &  -1.2486   &  0.006$\pm$0.010  &  0.105$\pm$0.016  &  0.025$\pm$0.006  &   0.060$\pm$0.006  &   0.038$\pm$0.006  &   0.019$\pm$0.007  &  0.036$\pm$0.003  &  0.031$\pm$0.007& \\
GC13   &  -1.155$\pm$0.252   &  0.014$\pm$0.009  &  0.118$\pm$0.022  &  --  &   0.089$\pm$0.008  &   0.072$\pm$0.012  &   0.043$\pm$0.007  &  0.040$\pm$0.005  &  0.022$\pm$0.004& \\ 
GC16   &  -0.542$\pm$0.417   &  0.064$\pm$0.018  &  0.138$\pm$0.041  &  0.057$\pm$0.009  &   0.172$\pm$0.018  &   0.136$\pm$0.019  &   0.066$\pm$0.015  &  0.049$\pm$0.011  &  0.059$\pm$0.007& \\  
GC27   &  -0.689$\pm$0.359   &  0.005$\pm$0.017  &  0.208$\pm$0.044  &  0.066$\pm$0.008  &   0.184$\pm$0.022  &   0.134$\pm$0.014  &   0.058$\pm$0.010  &  0.040$\pm$0.010  &  0.051$\pm$0.005& \\ %
GC29$^\dagger$   &  0.0660   &  -0.075$\pm$0.014  &  0.009$\pm$0.009  &  0.027$\pm$0.006  &   0.067$\pm$0.008  &   0.012$\pm$0.009  &   0.012$\pm$0.004  &  0.005$\pm$0.005  &  0.022$\pm$0.004& \\ 
GC44$^\dagger$   &  0.2883   &  0.022$\pm$0.023  &  0.005$\pm$0.036  &  0.107$\pm$0.017  &   0.227$\pm$0.022  &   0.130$\pm$0.016  &   0.075$\pm$0.019  &  0.103$\pm$0.015  &  0.072$\pm$0.008& \\ 
GC46   &  -0.175$\pm$0.379   &  0.122$\pm$0.028  &  0.232$\pm$0.046  &  0.090$\pm$0.015  &   0.235$\pm$0.022  &   0.176$\pm$0.021  &   0.086$\pm$0.015  &  0.089$\pm$0.012  &  0.070$\pm$0.008& \\ 
GC47$^\dagger$   &  -1.6469   &  -0.068$\pm$0.014  &  0.122$\pm$0.015  &  0.019$\pm$0.007  &   0.043$\pm$0.006  &   0.017$\pm$0.012  &   0.018$\pm$0.005  &  0.008$\pm$0.004  &  0.044$\pm$0.005& \\
GC59   &  -0.833$\pm$0.458   &  -0.034$\pm$0.011  &  0.121$\pm$0.016  &  0.066$\pm$0.007  &   0.126$\pm$0.010  &   0.083$\pm$0.014  &   0.057$\pm$0.007  &  0.027$\pm$0.006  &  0.169$\pm$0.033& \\ 
GC60$^\dagger$   &  0.2883   &  0.096$\pm$0.019  &  0.173$\pm$0.038  &  0.100$\pm$0.010  &   0.207$\pm$0.013  &   0.140$\pm$0.023  &   0.084$\pm$0.017  &  0.068$\pm$0.009  &  0.081$\pm$0.009& \\ 
GC61   &  -0.663$\pm$0.490   &  0.027$\pm$0.032  &  0.101$\pm$0.034  &  0.038$\pm$0.016  &   0.165$\pm$0.019  &   0.119$\pm$0.015  &   0.071$\pm$0.013  &  0.059$\pm$0.017  &  0.043$\pm$0.009& \\ 
GC62   &  -1.012$\pm$0.532   &  0.016$\pm$0.037  &  -0.092$\pm$0.041  &  0.035$\pm$0.013  &   0.093$\pm$0.022  &   0.028$\pm$0.013  &   0.050$\pm$0.014  &  0.033$\pm$0.015  &  -0.028$\pm$0.015& \\
GC66   &  -0.581$\pm$0.419   &  0.063$\pm$0.014  &  0.193$\pm$0.039  &  0.119$\pm$0.011  &   0.243$\pm$0.016  &   0.153$\pm$0.019  &   0.075$\pm$0.016  &  0.062$\pm$0.011  &  0.048$\pm$0.008& \\ 
GC70   &  -0.512$\pm$0.252   &  0.039$\pm$0.009  &  0.176$\pm$0.027  &  0.071$\pm$0.007  &   0.181$\pm$0.013  &   0.126$\pm$0.013  &   0.072$\pm$0.009  &  0.065$\pm$0.008  &  0.057$\pm$0.003& \\ 
GC79   &  -0.675$\pm$0.787   &  0.017$\pm$0.038  &  0.102$\pm$0.023  &  0.027$\pm$0.022  &   0.057$\pm$0.023  &   0.116$\pm$0.018  &   0.052$\pm$0.026  &  0.044$\pm$0.015  &  0.061$\pm$0.019& \\  
GC80   &  -0.735$\pm$0.239   &  -0.019$\pm$0.009  &  0.092$\pm$0.019  &  0.136$\pm$0.005  &   0.175$\pm$0.009  &   0.083$\pm$0.007  &   0.056$\pm$0.008  &  0.041$\pm$0.006  &  0.050$\pm$0.004& \\ 
GC87$^\dagger$   &  -1.2486   &  -0.017$\pm$0.025  &  0.019$\pm$0.027  &  --  &   0.019$\pm$0.011  &   0.030$\pm$0.009  &   0.068$\pm$0.028  &  0.008$\pm$0.011  &  0.022$\pm$0.010& \\
GC103  &  -0.914$\pm$0.283   &  0.230$\pm$0.019  &  0.118$\pm$0.016  &  0.033$\pm$0.005  &   0.138$\pm$0.009  &   0.085$\pm$0.013  &   0.061$\pm$0.011  &  0.050$\pm$0.006  &  0.035$\pm$0.003& \\ 
GC104  &  -0.490$\pm$0.255   &  0.048$\pm$0.008  &  0.185$\pm$0.033  &  0.083$\pm$0.006  &   0.195$\pm$0.014  &   0.133$\pm$0.011  &   0.083$\pm$0.009  &  0.065$\pm$0.009  &  0.055$\pm$0.004& \\ 
GC114$^\dagger$  &  0.2883   &  -0.021$\pm$0.011  &  0.150$\pm$0.028  &  0.127$\pm$0.007  &   0.182$\pm$0.014  &   0.114$\pm$0.007  &   0.066$\pm$0.007  &  0.052$\pm$0.009  &  0.058$\pm$0.005& \\
GC135  &  -0.934$\pm$0.258   &  -0.028$\pm$0.011  &  0.093$\pm$0.018  &  0.019$\pm$0.006  &   0.059$\pm$0.007  &   0.056$\pm$0.011  &   0.044$\pm$0.009  &  0.041$\pm$0.005  &  0.039$\pm$0.003& \\   %
GC136$^\dagger$  &   0.0660  &  0.095$\pm$0.016  &  0.212$\pm$0.024  &  -0.006$\pm$0.016  &   0.099$\pm$0.025  &   0.011$\pm$0.026  &   0.137$\pm$0.024  &  0.077$\pm$0.016  &  0.020$\pm$0.009& \\  
GC159  &  -0.819$\pm$0.271   &  0.034$\pm$0.005  &  0.148$\pm$0.020  &  0.078$\pm$0.004  &   0.199$\pm$0.013  &   0.094$\pm$0.006  &   0.058$\pm$0.006  &  0.034$\pm$0.008  &  0.044$\pm$0.004 & \\  
\hline
        \end{tabular}
           \begin{tablenotes}
                \begin{small}
                \item {\it Notes:} 
                (1) GC name. 
                (2) Measured metallicity.
                (3) CNR index. 
                (4) G index.
                (5) MgH index.
                (6) Mg2 index.
                (7) Mgb index.
                (8) Fe52 index.
                (9) Fe53 index.
                (10) Fe54 index.
                $\dagger$SSP model metallicity
             from the grid-method (Figure \ref{figura:hbeta_mgfe}).
                \end{small}
           \end{tablenotes}
        \end{scriptsize}
\end{table*}
\end{center}
\end{center}

\subsection{H$\beta$$^{\dagger}$ vs [MgFe]$'$ grid method}
\label{seccion:hbeta_mgfe}

To begin with, we follow the classical grid method suggested by \citet{Thomas:2004}  \citep[see also][]{BrodieHuchra1990} in the H$\beta$$^{\dagger}$ vs [MgFe]$'$  for a first approximation of age and metallicity. This involves measurement of hydrogen (H${\beta}$$^{\dagger}$), magnesium (Mgb$^{\dagger}$) and iron (Fe5270$^{\dagger}$ and Fe5335$^{\dagger}$) line strengths which were measured using the Lick/IDS index definitions from \citet{Trager:1998} for equivalent width:

    \begin{equation}
        EW = \int_{\lambda_{1}}^{\lambda_{2}} \left(1-\frac{F_{I}}{F_{BC}+F_{RC}}\right)d\lambda ,
        \label{equ:indice}
    \end{equation}
\noindent where $F_{I}$ is the mean flux measured in the center of the index (column (3) in Table \ref{tabla:indices_lino}), 
$F_{BC}$ and $F_{RC}$ are the mean flux in the continuum bandpass 
in both sides of the index, blue continuum (BC, column (2) in Table \ref{tabla:indices_lino}) and 
red continuum (RC, column (4) in Table \ref{tabla:indices_lino}).
The spectral ranges of the indices measured with this definition are listed in the last four rows of Table \ref{tabla:indices_lino}.
\citet{Thomas:2003} realized that the [MgFe]$'$ is independent of [$\alpha$/Fe] and is a good tracer of metallicity. This index is determined as: 
\begin{equation}
    [\rm MgFe]' = \sqrt{Mgb^{\dagger}(0.72\times Fe52^{\dagger} + 0.28\times Fe53^{\dagger})}.
    \label{equ:mgb_fe}
\end{equation}
\noindent We used those indices for constructing the index-index diagram for a first-approximation age and metallicity values. 
The calculated indices for all our GC candidates are tabulated in Table \ref{tabla:ancho_equivalente}. 
Error on each index was calculated by performing 1000
Monte Carlo simulations using the rms errors on the fluxes as the sigma of the Gaussian error distribution.
In Figure~\ref{figura:hbeta_mgfe}, we show our sample of GC candidates in the H${\beta}^{\dagger}$ vs [MgFe]$'$ diagram, where we also plot the theoretical age-metallicity grids using the SSP models of \citet{Bruzual:2003}. 
Most of the GC candidates in the figure have H${\beta}^{\dagger}<2.5$~\AA, and are located in the grid where the SSP ages are older than 8~Gyr and metallicity sub-solar,  
similar to the values for the Galactic GCs. Hence, for these objects, we obtained the metallicities using the empirical relation between the Fe index and metallicity suitable for the Galactic GCs, as explained in the next subsection.
However, there are 8 GC candidates whose location in Figure~\ref{figura:hbeta_mgfe} suggests that they are relatively young as compared to the classical Galactic GCs to be able to apply the Galactic Fe index-metallicity relation. 
Six of these have 
H${\beta}^{\dagger}\geq2.5$~\AA, suggesting ages less than $\sim$3~Gyr. The remaining  two objects (GC29, GC136) show H${\alpha}$ in emission, which is most likely originating from the diffuse ionized  gas in the disk, suggesting that the H$\beta^{\dagger}$ line is contaminated by the nebular line and hence for these the determined H$\beta^{\dagger}$ index is a lower limit, and hence likely to be younger than classical  Galactic GCs. For these eight objects, we assign the metallicity value of the closest SSP models.

Three of these 
(GC44, GC60, GC114)
have [MgFe]$'\geq2.5$~\AA, which corresponds to super-solar metallicity SSP models ([Fe/H]=0.288). The remaining GCs are closest to SSP tracks corresponding to  [Fe/H]=$-$1.249 (GC11, GC87), $-$1.647 (GC47) and 0.066 (GC29, GC 136).
The SSP metallicity obtained from the grid method is expected to be less accurate than that obtained using the Fe indices described below. Hence, the eight objects to which we assign their metallicity using the grid method are distinguished from the rest by a separate symbol in the corresponding Figures,  and are also marked by a dagger in Tables~\ref{tabla:indices_medidos} and \ref{tabla:resultados}.

\begin{figure} 
    \includegraphics[width=1.0\columnwidth]{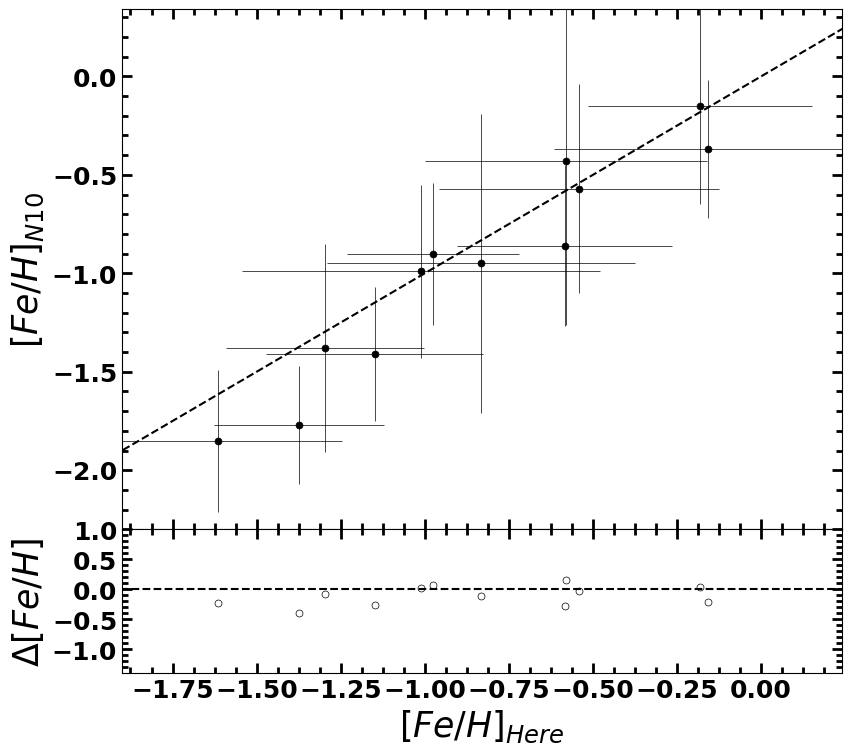}
    \caption{Metallicity comparison for M81 GCs. In the X-axis we plot our M81-GCs metallicities estimation, whereas in Y-axis are plotted the GCs metallicities estimation from N10. 
    }
    \label{figura:m81_comparation} 
\end{figure}

\begin{figure*} 
    \includegraphics[width=1.5\columnwidth]{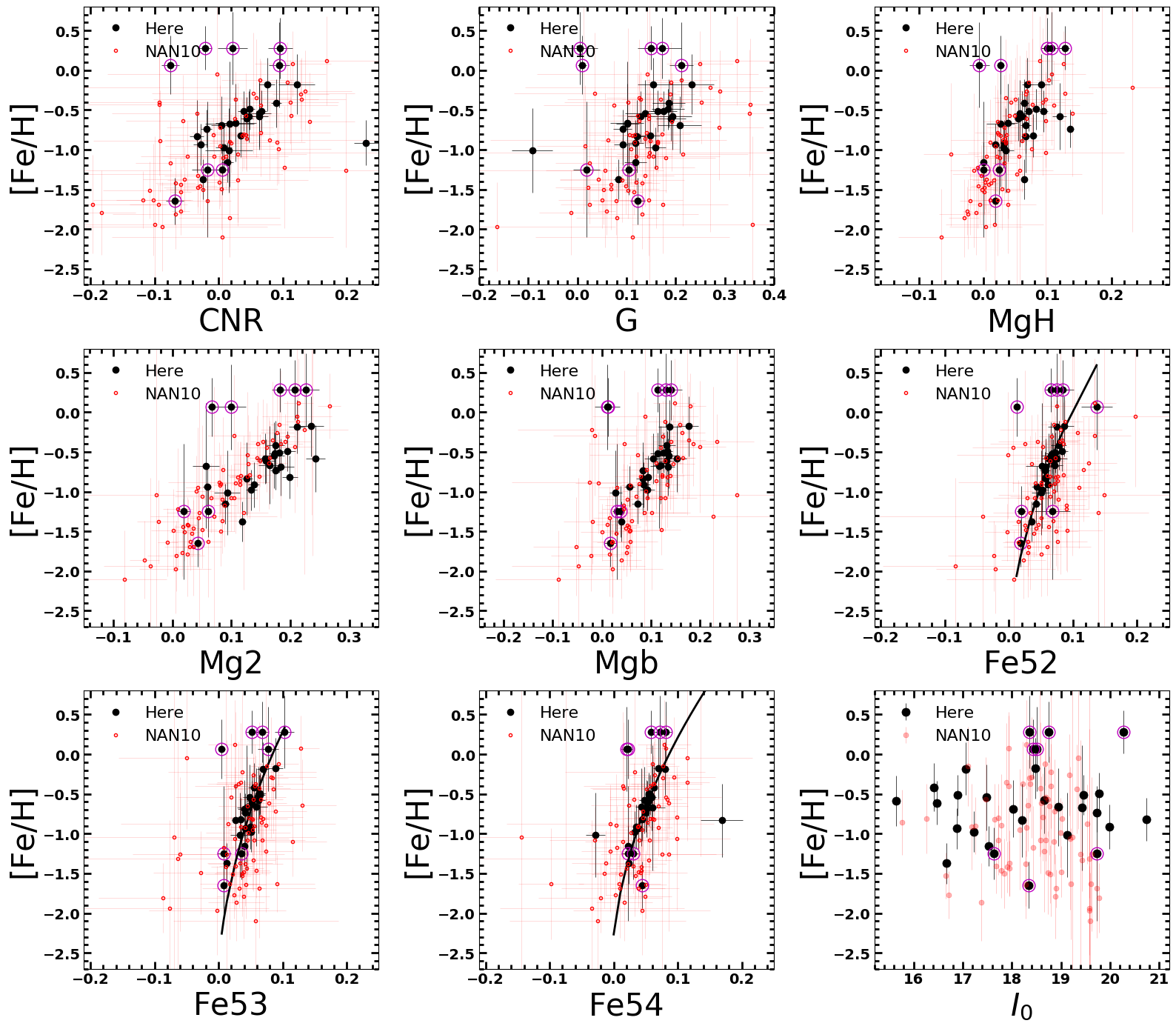}    
    \caption{\meta\ vs spectral indices. 
    {\it Black dots:} metallicity and spectral indices for each GC candidate. 
    {\it Red diffuse dots:} metallicity and spectral indices reported in N10.
    In the iron (Fe52, Fe53 and Fe54 vs \meta) windows the black solid lines are the curves tracing Equation~\ref{equ:meta_lino}.
    In last panel we show the \meta\ vs I-band magnitude.
    In all the panels the GCs whose metallicity was estimated using the grid method from Figure~\ref{figura:hbeta_mgfe} are encircled in magenta.}
    \label{figura:indices_medidos} 
\end{figure*}

\subsection{Metallicity from Fe indices}
\label{seccion:medidicon_metalicidad}

We calculated the well-known Lick indices \citep[][]{Burstein:1984} for all the GCs that have quality parameter G in Table~\ref{tabla:muestra}. 
For the calculation of spectral indices of our sample of GC candidates observed with GTC/OSIRIS (R1000B), we used the
definition from \citet{BrodieHuchra1990}:
    \begin{equation}
        I = -2.5\log_{10}\left[2\int_{\lambda_{1}}^{\lambda_{2}}\frac{F_{I}}{F_{BC}+F_{RC}}d\lambda\right] ,
        \label{equ:indice}
    \end{equation}

\noindent where $F_{I}$ is the mean flux measured in the center of the index (column (3) in Table \ref{tabla:indices_lino}), 
$F_{BC}$ and $F_{RC}$ are the mean flux in the continuum bandpass 
in both sides of the index, blue continuum (BC, column (2) in Table \ref{tabla:indices_lino}) and 
red continuum (RC, column (4) in Table \ref{tabla:indices_lino}).
The first eight indices listed in Table \ref{tabla:indices_lino} are measured using Eq.\,\ref{equ:indice}, 
whereas the remaining four are measured according to the definition of \citet{Trager:1998}.
Error on each index was calculated performing 1000 Monte Carlo simulations using Eq.\,\ref{equ:indice}, by taking into account the uncertainties of the fluxes $F_{I}$, $F_{BC}$ and $F_{RC}$. The calculated spectral indices, along with their errors, are listed in Table \ref{tabla:indices_medidos}.

We used the empirical relations between three iron indices (Fe52, Fe53 and Fe54) and the metallicity used in \citet{Mayya:2013} to infer the chemical abundance of the GC candidates. The behaviour of each spectral index as a function of the metallicity is well described by a second order equation:
    \begin{equation}
        ([{\rm Fe}/{\rm H}]-{\rm k})^{2} = 4{\rm p}({\rm Index}-{\rm h}).
        \label{equ:meta_lino}
    \end{equation}
\noindent where k, p and h are empirically derived coefficients.
The fits are displayed in the windows: Fe52, Fe53 and Fe54 in Figure~\ref{figura:indices_medidos}.
To calibrate the above relationship, the MW-GCs \citep{Schiavon:2005} and SSP models were used.
Because the SSP spectrum (see Section~\ref{seccion:spectral_fit} for SSP description) have a higher resolution than the observed, the indices and metallicities determination were obtained after degrading the model 
to the same wavelength step of the observed spectrum. Once our observed and SSP spectra were matched in resolution  and sampling, we used the Eq.\,\ref{equ:meta_lino} (coefficients k, p and h are given in Table \ref{tabla:lino_valores}) to calculate the metallicity. The metallicity reported in Table \ref{tabla:indices_medidos} is the weighted mean from the three iron indices. The error in \meta\ reported in Table \ref{tabla:indices_medidos} was calculated propagating the error of the Fe indices, along with the uncertainty of the coefficients: k, p, and h.

\begin{table}   
\begin{center}
\caption{Parameters of the second order equation which was used to estimate the 
metallicity.} 
\begin{tabular}{l c c c c }      
\hline
Index &   p   &     h &      k   \\  
\hline
\hline
 Fe52 &   28.39$\pm$0.24 & -0.004$\pm$0.0 & -3.41$\pm$0.02 \\ 
 Fe53 &   30.12$\pm$0.27 & -0.003$\pm$0.0 & -3.24$\pm$0.02 \\ 
 Fe54 &   23.93$\pm$0.16 & -0.005$\pm$0.0 & -3.96$\pm$0.01 \\ 
\hline
\end{tabular}
\label{tabla:lino_valores} 
\end{center} 
\end{table}

In Figure \ref{figura:m81_comparation}, we compare the metallicity obtained here with those determined by \citet{Nantais:2010} (hereafter N10) for 12 GC candidates that are common between the two samples. The two measurements agree between each other (solid line of unit slope), within the quoted errors. However, we note here that N10 reported a metallicity using the indices even for some of the candidates (eg. GC87) for which we have measured H$\beta$ index$>$2.5. As noted in the previous section, these GC candidates are younger than 5~Gyr and the index-metallicity relations are not reliable.

The \meta\ values measured by us as the mean of the Fe indices are plotted against all the eight indices (CNR, G, MgH, Mg2, Mgb, Fe52, F53, and F54) in Figure\,\ref{figura:indices_medidos} (black dots). We also show in this plot the corresponding values reported by N10 as {\it diffuse red dots}. Both the datasets show the expected correlation, with our measurements having, in general smaller spread, as compared to that of N10.
Also in last panel of Figure \ref{figura:indices_medidos}, we show the \meta\ vs $I$-band magnitude, from which it can be inferred that we have \meta\ measurements for candidates that are  1~magnitude fainter, and the dispersion is smaller over the entire range of magnitudes as compared to that of N10.

\subsection{Spectral fit: age and extinction determination}
\label{seccion:spectral_fit}

\begin{table} 
        \setlength\tabcolsep{5.5pt} 
	\centering  
         \begin{scriptsize}
	\caption{Fit parameters: age, extinction and $\chi^2$. } 
	\label{tabla:resultados} 
	\begin{tabular}{l r r r r c  }  
	\hline 
GC    & [Fe/H]$_{\rm model}$ & Age  & Age$_{\rm range}$ &  A$_{v}$ &  $\chi^{2}$ \\ 
      &             & (Gyr)  & (Gyr)       &  (mag)   &     \\ 
(1)   &  (2)        &  (3)   &  (4)        &  (5)     &  (6) \\ 
	\hline 
GC1  &  -0.6392 &  12.25$_{-1.25}^{1.50}$  &  11.00-13.75 &  0.75$\pm$0.07  &  1.52  \\  
GC3 &  -1.2486 &  10.00$_{-0.50}^{0.50}$  &  9.50-10.50 &  0.56$\pm$0.07  &  3.32  \\  
GC4  &  -0.6392 &  12.75$_{-0.50}^{0.50}$  &  12.25-13.25 &  0.89$\pm$0.03  &  2.89  \\  
GC5  &  -0.3300 &  13.75$_{-0.25}^{0.00}$  &  13.50-13.75 &  0.10$\pm$0.03  &  1.43  \\  
GC7 &  -0.6392 &  1.50$_{0.50}^{0.50}$  &  1.00-2.00 &  0.80$\pm$0.18  &  2.17  \\  
GC9 &  -0.3300 &  11.50$_{-1.75}^{1.75}$  &  9.75-13.25 &  0.50$\pm$0.18  &  1.31  \\  
GC10 &  -1.2486 &  13.75$_{-0.25}^{0.00}$  &  13.50-13.75 &  0.89$\pm$0.10  &  3.42  \\  
GC11$^{\dagger}$ &  -1.2486 &  12.25  &  2.00,12.25$^{\ddagger}$&  0.29$\pm$0.03  &  1.99 \\  
GC13 &  -1.2486 &  13.75$_{-0.25}^{0.00}$  &  13.50-13.75 &  0.86$\pm$0.12  &  2.27  \\  
GC16 &  -0.6392 &  12.75$_{-0.25}^{1.0}$  &  12.50-13.75 &  1.32$\pm$0.04  &  2.53  \\  
GC27 &  -0.6392 &  12.75$_{-0.75}^{1.0}$  &  12.00-13.75 &  -  &  1.81  \\  
GC29$^{\dagger}$ &  0.0660 &  1.00$_{0}^{0.50}$  &  1.00-1.50 &  -  &  0.83  \\  
GC44$^{\dagger}$ &  0.2883 &  7.50$_{-2.00}^{2.50}$  &  5.50-10.00 &  0.34$\pm$0.19  &  1.60  \\  
GC46 &  -0.3300 &  13.75$_{-0.50}^{0.0}$  &  13.25-13.75 &  0.67$\pm$0.19  &  1.99  \\  
GC47$^{\dagger}$ &  -1.6469 &  6.50  & 1.00,6.50$^{\ddagger}$ &  1.79$\pm$0.10  &  2.33 \\  
GC59 &  -0.6392 &  3.00$_{-0.50}^{0.50}$  &  2.50-3.50 &  2.22$\pm$0.17  &  1.81  \\  
GC60$^{\dagger}$ &  0.2883 &  2.50$_{-0.50}^{0.50}$  &  2.00-3.00 &  0.37$\pm$0.12  &  1.20  \\  
GC61 &  -0.6392 &  12.75$_{-6.25}^{1.0}$  &  6.50-13.75 &  1.39$\pm$0.11  &  1.79  \\  
GC62 &  -1.2486 &  13.25$_{-0.75}^{0.5}$  &  12.50-13.75 &  1.11$\pm$0.18  &  2.16  \\  
GC66 &  -0.6392 &  12.75$_{-1.0}^{1.0}$  &  12.75-13.75 &  0.83$\pm$0.14  &  2.64  \\  
GC70 &  -0.6392 &  13.25$_{0.75}^{0.5}$  &  12.5-13.75 &  0.31$\pm$0.11  &  2.91  \\  
GC79 &  -0.6392 &  13.75$_{2.25}^{0.0}$  &  11.50-13.75 &  1.11$\pm$0.18  &  1.91  \\  
GC80 &  -0.6392 &  3.00$_{-0.50}^{0.50}$  &  2.50-3.50 &  0.26$\pm$0.15  &  1.30  \\  
GC87$^{\dagger}$ &  -1.2486 &  4.00  &  1.00,4.00$^{\ddagger}$ & 1.25$\pm$0.30 & 1.19  \\
GC103 &  0.0660 &  2.50$_{-0.50}^{0.50}$  &  2.00-3.00 &  0.43$\pm$0.26  &  1.82  \\  
GC104 &  -0.6392 &  13.25$_{-1.00}^{0.50}$  &  12.25-13.75 &  0.63$\pm$0.14  &  4.25  \\  
GC114$^{\dagger}$ &  0.2883 &  2.50$_{-0.50}^{0.50}$  &  2.00-3.00 &  0.43$\pm$0.19  &  1.61  \\  
GC135 &  -0.6392 &  2.00$_{-0.50}^{0.50}$  &  1.50-2.50 &  1.18$\pm$0.19  &  0.99  \\ 	
GC136$^{\dagger}$ &  0.0660 &  2.50$_{-0.5}^{0.5}$  &  2.00-3.00 &  0.48$\pm$0.14  &  1.77  \\ 
GC159 &  -0.6392 &  6.50$_{-0.5}^{0.5}$  &  6.00-7.00 &  0.64$\pm$0.07  &  3.00  \\  
\hline  
        \end{tabular}  
           \begin{tablenotes}  
            \begin{small}  
            \item {{\it Notes:} (1) GC candidates name.
            (2) Metallicity model.
            (3) Age and its error determined with the $\chi^{2}$.
            (4) Age range.
            (5) Visual extinction and its error. 
            (6) $\chi^{2}$ Minimization value.
            $\dagger$Metallicity
            model from the grid-method (Figure \ref{figura:hbeta_mgfe}).
            $\ddagger$Contains two populations. The two ages  correspond to that of the metal-rich (young) and metal-poor (old) populations that best fit the spectra, with the [Fe/H] in column 2 corresponding to that of the older of the two populations.} 
            \end{small}  
           \end{tablenotes}                  
        \end{scriptsize}
\end{table}    

We determined the age for each GC candidate by selecting an SSP spectrum that best fits our observed spectrum. The fitting is carried out using a $\chi^2$ procedure, which is explained below.
We used Charlot \& Bruzual models (in-preparation), which are the updated \citet{Bruzual:2003} models \citep[see][for details]{Plat:2019, Mayya:2023}. These models are 
based on MILES \citep[][]{Miles:2006} library synthesized for the Kroupa \citep[][]{Kroupa:2003} initial mass function, 
at six metallicities Z=0.0001, 0.0004, 0.004, 0.008,  0.02, and 0.05, and ages from 1 to 13.75~Gyr.
For a given GC candidate, we used the SSP model that has metallicity closest to the \meta\  value measured in the previous section.
These model spectra are smoothed to match the resolution of the observed spectra of $\sim$7~\AA. Both the observed and model spectra are then divided by average flux  in the wavelength  range from 5490 to 5510 \AA,
to get the normalized spectra. Subsequently, the spectra were interpolated linearly in order to have the same sampling in the observed $\lambda$ range. The $\chi^2$-fitting is carried out in intervals of $\Delta\lambda\sim300$ {\AA} over the spectral range $I_{\lambda}=[3650,5600]$\,{\AA}. The spectra redward of 5600\,\AA\ do not contain age-sensitive absorption features, which is the reason for excluding this part in our $\chi^2$ analysis. 
The piece-wise, rather than the entire spectral, fitting procedure ensures that the $\chi^2$ values are not affected by differences in the shapes of the observed and model spectra over scales larger than $\Delta\lambda$. Such differences are common due to the wavelength-dependent flux calibration errors, and/or due to residual errors in the subtraction of the sky in case of fainter objects. The $\chi^2$ values obtained in the adopted window size is only weakly dependent on the value of the extinction, which allows us to obtain $A_V$ in an iterative manner as explained later in this section.

Then, for a given SSP model, we compute the reduced $\chi^{2}_k$ value within each $k^{\rm th}$-window using the formula
\begin{equation}
    \chi^{2}_{k} = \frac{1}{\nu} \sum_{i=1}^{n_k}\left(\frac{{\rm F}_{{\rm obs},k}(\lambda_i) - {\rm F}_{{\rm mod},k}(\lambda_i) }{\sigma_{{\rm obs},k}}\right)^{2}, 
    \label{equ:chi2}
\end{equation}

\noindent where $n_{k}$ is the number of spectral points with well-measured fluxes,  ${\rm F}_{{\rm obs},k}$ is the observed spectrum, ${\rm F}_{{\rm mod},k}$ is the model spectrum,
$\sigma_{{\rm obs},k}$ is the standard deviation of the flux  
and $\nu=n_k-1$ is the number of degrees of freedom.
We defined masks to ignore spectral regions that are affected by the subtraction of bright sky lines.
The value of $n_{k}$ is the number of unmasked pixels within the window. The $\sigma_{{\rm obs},k}$ is given by
    \begin{equation}
        \sigma_{{\rm obs},k}=\sqrt{  \frac{1}{n_k-1}\sum_{i=1}^{n_k} \left({\rm F}_{{\rm obs},i}-\overline{\rm F}_{\rm obs}\right)^{2}  },
        \label{equ:sigma}
    \end{equation} 
where $\overline{\rm F}_{\rm obs}$ is the mean of observed flux in the $k^{\rm th}$ window. The value of $\sigma_{{\rm obs},k}$  is obtained iteratively, clipping the observed points that lie above and below 3$\sigma$ and re-calculating a new ${\text F}_{{\text obs},k}$, $\overline{\rm F}_{\rm obs}$ and $\sigma_{\rm obs,k}$ in each iteration. The $\sigma_{\rm obs,k}$ values obtained after $\sim$5 iterations do not change much, and hence we use the value after 5 iterations as the typical standard deviation of the spectrum in the window under analysis.

The final value of $\chi^{2}$ over the entire spectral range is calculated as the average of all $\chi^{2}_k$.
This process is repeated for all ages of the SSPs between 1 and 13.75~Gyr and the model age with the minimum $\chi^{2}$ is taken as the best age of the GC candidate.

For the three cases (GC11, GC47, GC87) that have EW(H$\beta$)$>$2.5~\AA\ and sub-solar metallicities, the residual clearly shows H$\beta$ in absorption suggesting a necessity for a second younger population. We hence found a new $\chi^2$ solution where the input SSP is a linear combination of an old, metal-poor and young, solar metallicity populations. The $\chi^2$ values for the combined SSP model were found to be lower than with only a single population for these three GCs.

We clarify that the extinction differences within the 300~\AA\ window used for obtaining $\chi^{2}_{k}$ are ignored in Eq.~\ref{equ:chi2}, which is justifiable given the relatively small wavelength interval. However, we take into account the extinction correction in an iterative way. The observed spectrum over the entire observed wavelength is systematically redder as compared to the best-fit model spectrum. We used the slope between $\lambda_{B}$=4400, the normalized wavelength, $\lambda_{V}$, and $\lambda_{R}$=6400~\AA, to obtain the $A_{V}$ using the following formula:

\begin{equation}
  A^{B,R}_{V}=\frac{\log (F^{B,R}_{\rm obs})- \log (F^{B,R}_{\rm mod})}{-0.4\cdot[E_{\rm MW}(\lambda^{B,R})-1]},
  \label{equ:extinction}
\end{equation} 

\noindent where the $F^{B}_{\rm obs}$, $F^{R}_{\rm obs}$, $F^{B}_{\rm mod}$ and $F^{R}_{\rm mod}$ are the average values over $\sim$200~\AA\ windows centered at $\lambda_B$ and $\lambda_R$ measured in the observed and best-fit model spectrum, respectively, and $E_{\rm MW}$ is the MW extinction curve from \citet{Cardelli:1989}.
The $A^B_{V}$ and $A^R_{V}$ values are averaged to get the mean $A_{V}$ and error in the corresponding $A_{V}$. All model spectra are reddened using the determined $A_{V}$, normalized to its flux at $\lambda_{V}$ and the entire fitting-procedure to obtain the minimum $\chi^{2}$ is repeated. The iterative procedure is stopped when the age and $A_{V}$ values in two consecutive iterations converge. We find that the solutions converge after 2 and 5 iterations for high and low SNR spectra, respectively.

The ages of the SSP model and extinction for the best-fit case are 
tabulated in Column 3 and 5 in Table \ref{tabla:resultados} respectively. 
According to the $\chi^2$-statistics all models that have $\nu(\chi^{2}-\chi^{2}_{\rm min})=1$ are acceptable solutions within 1$-\sigma$ with 1 parameter involved
\citep[][]{Wall:2012}.
We used this criterion to estimate the error on the best-fit ages. 
The possible range of ages, which are obtained by applying the lower and upper errors, are given in Column 4.

\begin{figure}
    \includegraphics[height=5.cm,width=8.5cm]{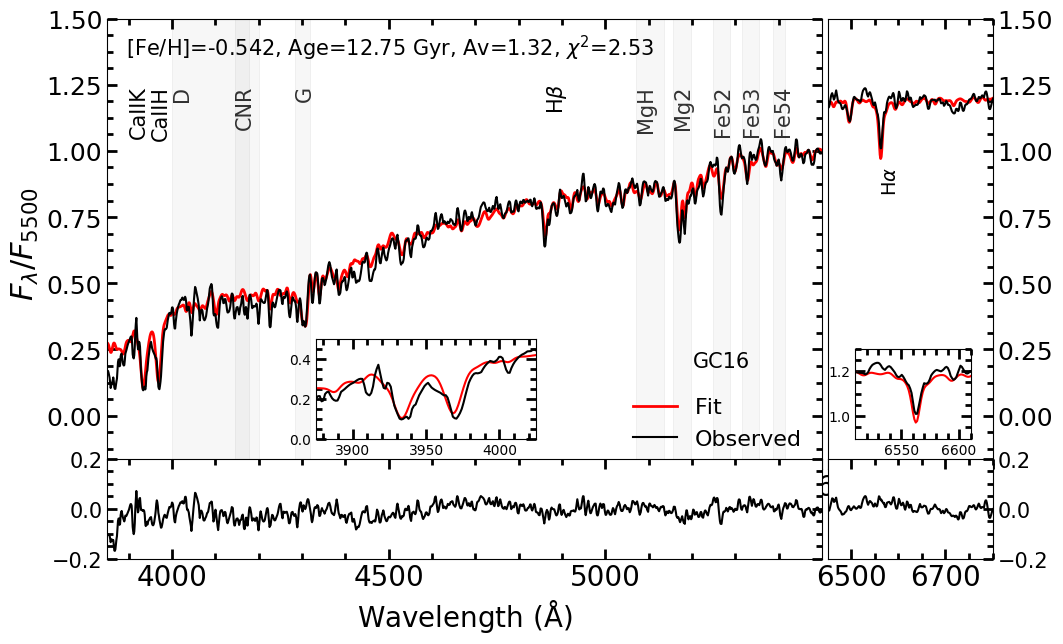}
    \caption{Observed spectrum (black line) and the spectral fit (red line).
    Gray bands indicate the wavelength coverage of the spectral indices. The fitted parameters are: \meta$=-0.54\pm0.10$, Age=12.75 Gyr,  $A_{V}$=1.32 and $\chi^2=2.53$.
    The bottom plots show the residuals between the observation and the model.
    Interior windows are zooming the regions of the CaH+K (left) and H$\alpha$ (right) lines.
    }
    \label{figura:espectro_observado_ajustado}
\end{figure}

\begin{figure} 
    \includegraphics[width=0.8\columnwidth]{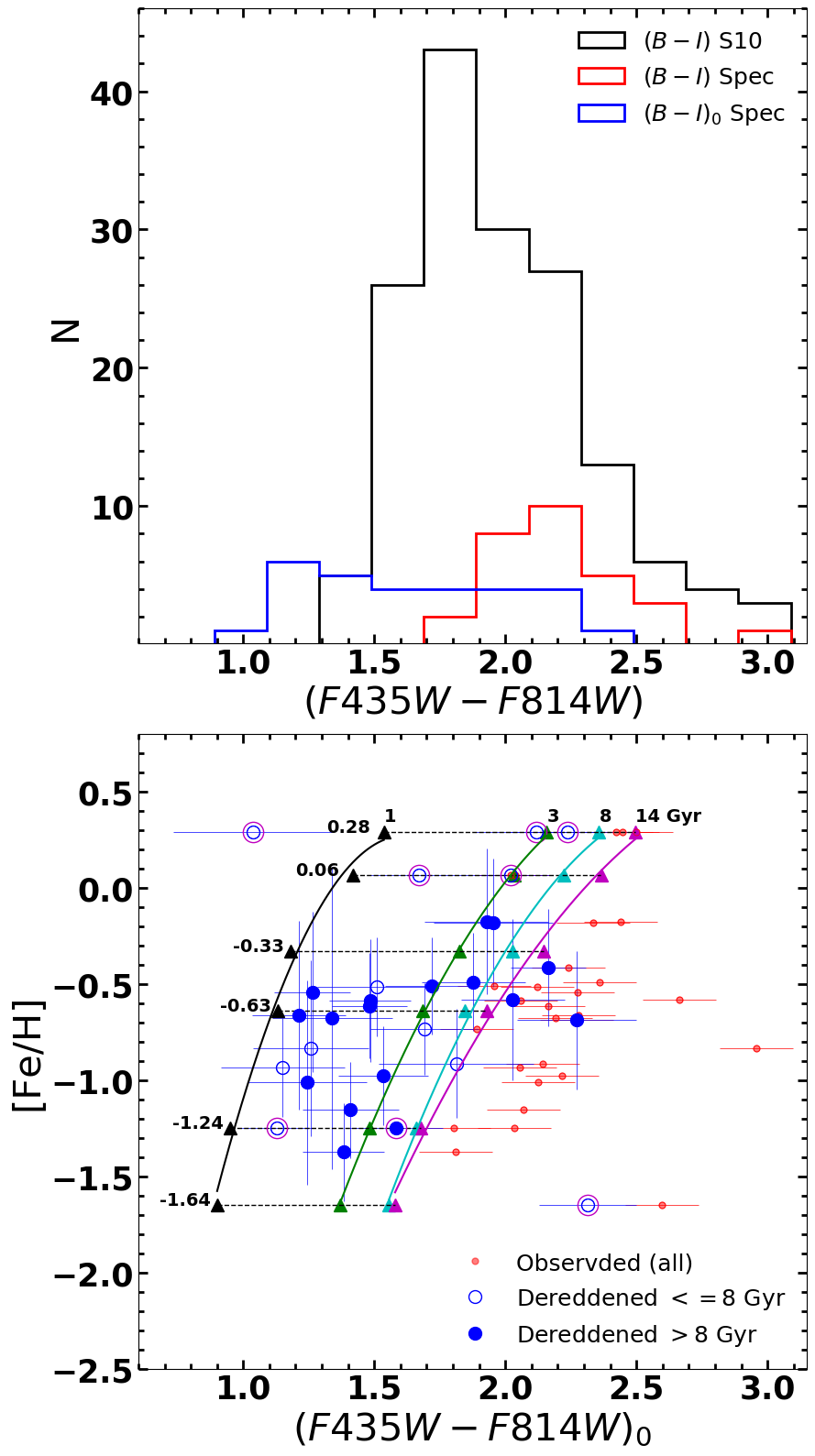} 
    \caption{{\it Top:} $(B-I)_{0}$ colour distribution. 
             {\it Black histogram:} whole sample from S10 corrected for the Galactic reddening.
             {\it Blue histogram:} sample of GC candidates corrected for the Galactic and intrinsic reddening. 
             {\it Red histogram:} sample of GC candidates corrected only for the Galactic reddening.
             {\it Bottom:} metallicity-colour relation for the sample of GC candidates with spectroscopic observations after ({\it blue circles}) and before ({\it red circles}) intrinsic reddening correction. Also the {\it blue circles} are coded by their spectroscopic ages as indicated by the legends at the bottom right corner.
            SSP model grids from \citet{Bruzual:2003} with Kroupa IMF between 0.1 and 100~\mso\ are over-plotted for four fixed ages between 1 and 14~Gyr and six metallicities between $-1.64\leq \meta\leq +0.28$.
            The solid coloured lines connecting  grid points at fixed age are 
            second order polynomials fitted to the SSP models.
            The GCs whose metallicity was estimated using the grid method from Figure~\ref{figura:hbeta_mgfe} are encircled in magenta.
            }            
\label{figura:metalicidad_color} 
\end{figure}

In Figure \ref{figura:espectro_observado_ajustado} we show the fitted spectrum for one of the representative GCs, GC16. The measured \meta$=-0.54\pm 0.08$ for which we used the \meta$_{model}=-1.2486$. The best-fit age=12.75$_{-0.25}^{+1.0}$~Gyr and $A_{V}$=1.32$\pm0.02$. The residual is shown in the bottom panel. In order to illustrate the goodness of the fit, we zoom in on the Ca H+K feature, and the H$\alpha$. It can be appreciated that the best-fit SSP for the determined $A_V$ reproduces the observed strength of the H$\alpha$ line as well as the continuum level very well, in spite of the fact that the fitting was carried out blueward of 5600~\AA.


\section{Discussion}

Our spectroscopic sample consists of 42 GC candidates, among which 30 have good quality spectra for a reliable determination of their metallicities, ages and extinctions. 
Spectra of all the 42 however are good enough for the measurement of recessional velocity (see column 7 in Table~\ref{tabla:muestra}). Being  a sub-sample of the S10 sample, all also have well-measured colours and magnitudes. Given that GC populations of very few spiral galaxies have all these data available, we use these data to both characterise the properties of GC samples as well as to address the origin of GCs in a Cosmological context.
In the first two sections, we discuss colour-metallicity relation and age distribution of our spectroscopic sample.
In section~\ref{seccion:spatial_distribution}, we combine the velocity of our sample objects with that from N10 to explore whether the majority show kinematics expected for halo or disk objects. In section~\ref{seccion:metallicity_age}, we combine our age and metallicity data with that from other samples that are analysed using methods similar to ours to discuss GCs from a Cosmological perspective.

\subsection{Metallicity-colour relation}
\label{seccion:metallicity_colour}

\begin{table*}
\begin{center}
        \setlength\tabcolsep{4.8pt}
	\caption{Results from GMM and dip test.}
	\label{tabla:bimodalidad_ssc_gcs}
	\begin{tabular}{l c c c c c c c c c c c c c l c} 
\hline
Colour   & Peak1 &  $\sigma1$&   Peak2 & $\sigma2$&   \ngc & $f_{2}$ &    D &  \multicolumn{3}{c}{$p$-values}  &  Kurtosis &   Dip &   Bi \\
(1)&        (2)  &   (3)  &   (4)   &      (5) &    (6)&     (7)&   (8) &  \multicolumn{3}{c}{(9)} &   (10)  & (11) \\
\hline
$(B-I)_{0}$ & 2.15$\pm$0.11 & 0.24$\pm$0.07 & 2.64$\pm$0.31 & 0.42$\pm$0.11 & 153 & 0.19$\pm$0.26 & 1.42$\pm$1.09 & 0.01 & 0.68 & 1.0 &  2.48  & 0.11  & N \\ 
\hline

	\end{tabular}
           \begin{tablenotes}
                \begin{small}
    \item {\it Notes:} (1) Galaxy. 
    (2-3) Mean and standard deviation of the first peak in the double-Gaussian model. 
    (4-5) Mean and sigma of the second peak in the double-Gaussian model. 
    (6) Total number of GCs. 
    (7) Fration of \ngc\ associated with the second peak. 
    (8) Separation of the means relative to their widths. 
    (9) GMM $p$-values based on the likelihood-ratio test $p$($\chi^{2}$), 
    peak separation $p$(DD), and 
    kurtosis $p$(kurt) (lower $p$-values are more significant).
    (10) Kurtosis of the colours distribution. 
    (11) Dip value is the significance level with which a unimodal distribution can be rejected. 
    (12) Bimodality final evaluation: Y (Yes), N(No).
    We note that it was only possible to fit 153 of the 172 GC candidates reported in S10.
                \end{small}
            \end{tablenotes}
\end{center}
\end{table*}

One of the most frequently discussed topics in the study of GCs is the bimodal colour distribution, which is thought to be originated from a metallicity-colour relation \citep[see e.g.][]{Brodie:2006}. 
It may be recalled that our spectroscopic sample is drawn from S10, which is a photometric sample of 172 GC candidates covering the entire FoV offered by the HST/ACS footprints displayed in Figure~\ref{figura:footprint}. Our spectroscopic sample is not a representative subset of the original sample in colour and is relatively of small size, which prevents us an exploration of the issue of bimodality in metallicity. Besides, the original S10 (black solid histogram in top panel of Figure~\ref{figura:metalicidad_color}) sample itself does not demonstrate an obvious colour bimodality. We confirmed an absence of evidence of colour-bimodality in the S10 sample  using the Gaussian Mixture Modeling (GMM) code \citep{Muratov:2010} which carries out a robust statistical test for evaluating bimodality, and uses the likelihood-ratios to compare the goodness of the fit for double-Gaussian versus a single-Gaussian. This method is independent of the binning of the sample. The results from the GMM test in the colour distribution are shown in Table \ref{tabla:bimodalidad_ssc_gcs}
(see the footnote for the explanation of the column headers). For a distribution to be considered bimodal, the Kurtosis must be negative, D$>2$ and $p$-values must be small. We found a Kurtosis of 2.48  and a separation (D) of 1.42, values that strongly discard bimodality in the distribution.

We also used ``dip test" \citep{Hartigan:1985}, which calculates the maximum distance between the cumulative input distribution and the best-fitting unimodal distribution. This test estimates the significance level with which a unimodal distribution can be rejected.
The probability given by the ``dip test" to be a unimodal distribution is of 89\%.
After carrying out these two independent statistical tests, we conclude that the M81 GCs of the S10 sample do not show evidence for bimodality in colour. This may not necessarily mean an absence of bimodality among all M81 GCs, as the S10 sample is more biased towards the disk GCs and lacks halo GCs, which are known to be systematically metal-poor and blue, as compared to the disk GCs.

Several studies have addressed the issue of 
a relation between metallicity and colour of GCs from both observational (e.g. \citealt{Larsen:2001, Brito:2011, Fahrion:2020, Kim:2021}) and theoretical (e.g. \citealt{Yoon:2006, Cantiello2:2007}) perspectives. The availability of $A_V$ intrinsic to the objects in our spectroscopic sample allows investigating this relation with the reddening-corrected colours. The latter correction was carried out using the $A_V$ in column~5 of Table~\ref{tabla:resultados} and the  \citet{Cardelli:1989} reddening curve. 
In the bottom panel of Figure \ref{figura:metalicidad_color}, we show the metallicity-colour relation for the spectroscopic sample, before (red circles) and after (blue circles) reddening  correction. The colour distribution of the spectroscopic sample before and after reddening correction is shown in the top panel. 
We also show the theoretical age-metallicity grids using the SSP models of 1, 3, 8 and 14 Gyr (black, green, cyan and magenta triangles, respectively) from \citet{Bruzual:2003}, with Kroupa IMF (from 0.1 to 100 $\mathcal M_{\odot}$), and six metallicities: Z=0.0004, 0.001, 0.004, 0.008, 0.019 and 0.030. 
At a given age, the metallicities and SSP model colours are well-fitted by second order polynomial curves, which are shown by solid lines. The SSP model curves for the two extreme ages (1 and 14~Gyr) enclose the  distribution of observed points after dereddening the colours. 
Note that the observed colours before reddening correction for all objects are redder than the values expected for the oldest SSPs.
We find a mean extinction of $A_V=0.80\pm0.13$~mag for our spectroscopic sample of GCs in M81. This relatively high values illustrate the importance of reddening corrections before any robust statistical analysis of colours, such as the exploration of bimodality, colour-metallicity relation etc.
 
It is interesting to note that the uncorrected (i.e. observed) colours of the majority of clusters of our spectroscopic sample are redder than the SSP values expected for their spectroscopic ages. The reddening correction, however, makes the colour bluer for their derived ages. Errors in both the axis (horizontal error bars take into account the errors on our $A_V$ measurements) can account for some of this inconsistency. It is also likely that a hidden younger second population is making the colours bluer for the old GCs (solid blue dots). A detailed multi-colour study would be required to simultaneously explain the metallicity, age and colours for each object, which is beyond the scope of the current discussion. Notwithstanding, the figure suffices to illustrate the existence of an empirical metallicity-colour relation.

\begin{figure}
    \includegraphics[width=0.9\columnwidth]{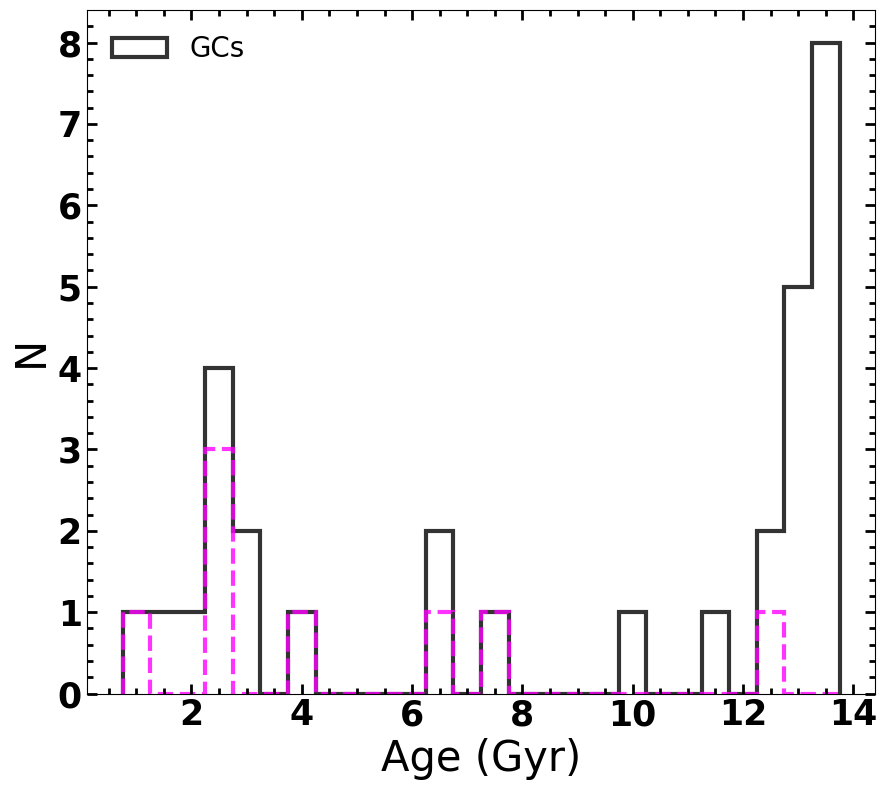}  
    \caption{Age distribution of GC candidates. The bulk of the sample has an old age typical of GC. The magenta histogram corresponds to GCs whose metallicity was estimated using the grid method from Figure~\ref{figura:hbeta_mgfe}.}
    \label{fig:edad_distribucion}
\end{figure}

\subsection{Age distribution of candidate GCs}
\label{seccion:age_discusion}

\begin{figure} %
\includegraphics[width=\columnwidth]{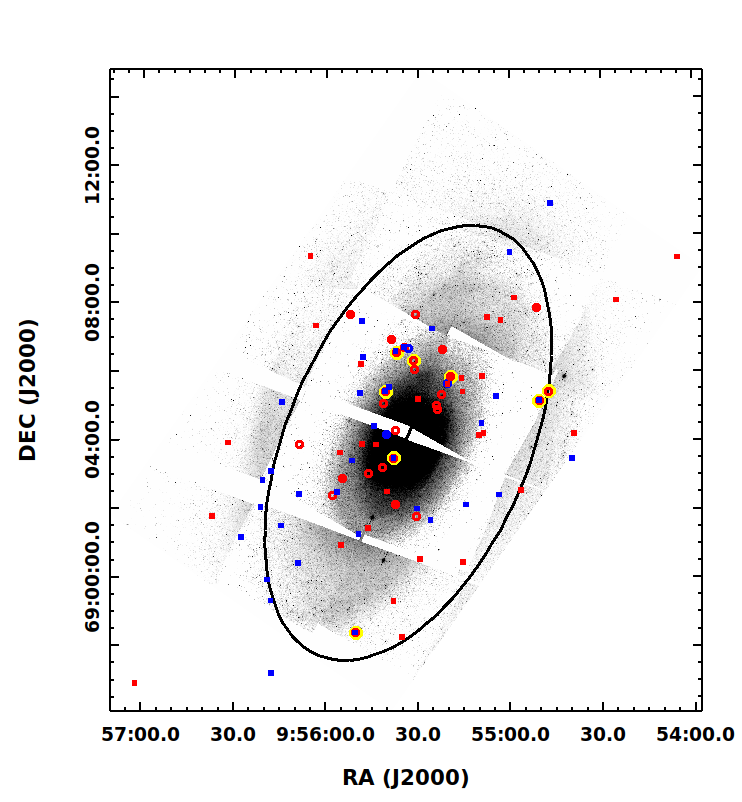}
    \caption{ Spatial distribution of GCs with measured spectra, superposed on a $F814W$-HST image. 
             {\it Blue points:} metal-poor GCs (\meta$\leq-$1.0). {\it Red points:} metal-rich GCs (\meta$>-$1.0).
             Black ellipse is 0.5 times $R_{25}$. We show our sample as big circles (6$\arcsec$)
             and the N10 sample as small squares (3$\arcsec$). The GCs whose metallicity was estimated using the grid method from Figure~\ref{figura:hbeta_mgfe} are encircled in yellow.             
             North up and east left.}
    \label{fig:dist_espacial}
\end{figure}

Ages of GC candidates were determined using the $\chi^2$ fitting between the observed and SSP spectra as discussed in section \ref{seccion:spectral_fit}.
In Figure \ref{fig:edad_distribucion}, we show the age distribution of our spectroscopic GC sample.
The distribution peaks at the oldest age bin, with 17 GCs being older than 8~Gyr. We consider these 17 cases as genuine ``classical GCs'' similar to the ones in the MW. The remaining 13 cases, including seven cases for which we estimated metallicity using the grid-method, have ages between 1 and 8~Gyr, which are clearly younger than the typical Galactic GCs. We refer to these 13 cases, 9 of which are younger than 3~Gyr, as ``intermediate age star clusters''.

\begin{figure*} 
    \includegraphics[width=0.9\columnwidth]{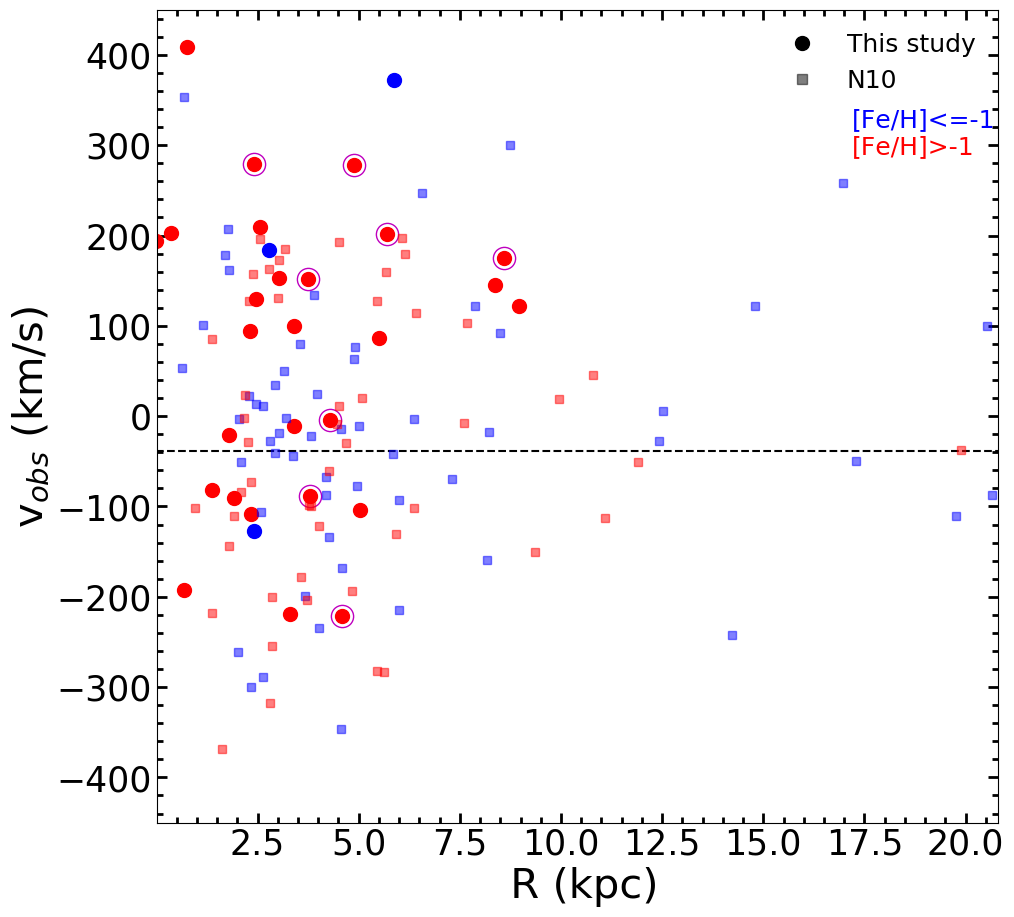}
    \includegraphics[width=0.97\columnwidth]{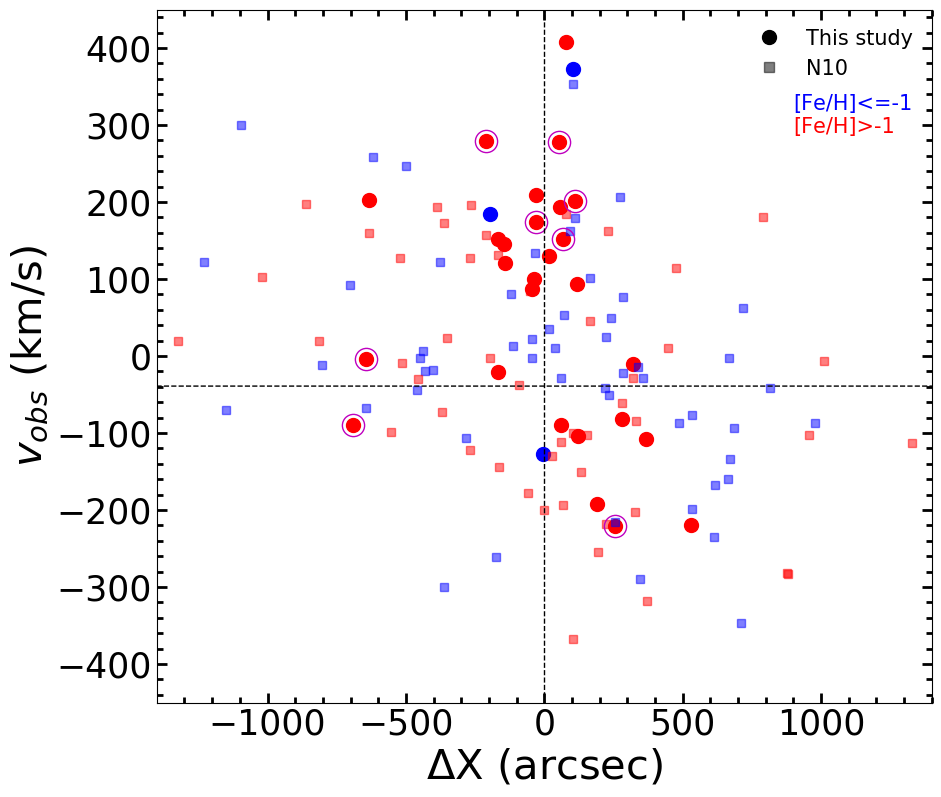}
    \caption{Phase-space diagram of the GC candidates, with
    metal-poor \meta\ $\leq-1.0$ ({\it blue color}) and metal-rich \meta\ $>-1.0$ ({\it red color}) distinguished. The observed velocity is plotted against galactocentric distance (left) and distance along the major axis (right).
    The horizontal dashed line in both panels is the galaxy systemic velocity. 
    In right panel the vertical line shows the centre of major axis. In both plots we show our spectroscopic sample as big circles and the N10 sample as small squares. Also in both plots GCs whose metallicity was estimated using the grid method from Figure~\ref{figura:hbeta_mgfe} are encircled in magenta.
    }
    \label{fig:velocities} 
\end{figure*}

Ages of our spectroscopic sample
suggests that as much as 43\% (13/30) of GC candidates are not classical GCs like those in the MW. The main reason for this relatively large contamination fraction is the age-extinction degeneracy which results in the reddened relatively younger clusters to have similar $F435W-F814W$ colours as that of unreddened classical GCs.
L22 used $u-g$ vs $F435W-F814W$ colour-colour diagram to break this degeneracy and determine
the contamination of GC candidate samples from reddened young clusters (age $<$3~Gyr) for a sample of five spiral galaxies, including M81. They found a contamination fraction of 32\% 
in M81. 
Our spectroscopic sample covered the very inner part of M81, where the extinction and contamination from younger populations are expected to be higher, which is the most likely reason for slightly  
higher contamination factor for the spectroscopic sample. 

\subsection{Velocities and spatial distribution} 
\label{seccion:spatial_distribution}

In early-type galaxies the distribution of metal-poor GCs extends to the outer halos while metal-rich ones are more concentrated to the internal parts
\citep[see e.g.][]{Hargis:2014, Kartha:2014}. 
However, this is not entirely proven for late-type galaxies, since studies of GCs in these kind of galaxies are scarce \citep[see e.g.][]{Hargis:2014}. 
Metal-poor outer halo GCs and metal-rich disk GCs also distinguish in their kinematical behaviour with the former being pressure supported, whereas the rotation dominates the kinematics of the latter. Testing these scenarios require GC samples with velocity and metallicity measurements that cover both halo and disk components. Unfortunately GCs in our spectroscopic sample reside in the inner parts of the galaxies, and lack halo GCs (for simplicity in our sample we called GCs to both intermediate age star clusters and classic GCs, unless a distinction is made between them). In order to overcome this bias, we complement the velocities and metallicities  for the 30 GC candidates of our spectroscopic sample with these quantities for 108 GCs from N10 sample, which contains the outer halo GCs. The combined sample allows us to examine whether the metal-poor and metal-rich GCs differ in their kinematical properties.

In Figure \ref{fig:dist_espacial}, we show the spatial distribution of GCs for the combined sample superposed on a HST/ACS $F814W$ image. 
The metal-poor and metal-rich GCs are shown in separate colours. We also show an ellipse with semi-major axes a=6.725\arcmin\ and b=3.525\arcmin\, which encloses the region defined by 0.5$R_{25}$ \citep{RC3}. 
Unfortunately, even the combined sample has very few GCs outside the ellipse, and hence the most complete sample that can be constructed today for M81 is skewed towards disk GCs. 

In left panel of Figure \ref{fig:velocities}, we show the 1D phase-space diagram from this work (big circles) and N10 (small squares) samples of GC candidates, where the galaxy systemic velocity, $-$38.97$\pm$2.69~km~s$^{-1}$ \citep{Speights:2012} is shown as a dashed line. Metal-poor and metal-rich GCs are shown in blue and red  colours, respectively. 
In such diagram, the object location is related to the kinematics of the component it belongs to (i.e. disk or bulge/halo) and to the time since first in-fall into the galaxy halo \citep{Rocha:2012}. In fact, the outer envelope in the $R$ vs $v_{obs}$ space is dominated by recently accreted objects, while objects with smaller difference to the galaxy systemic velocity and  lower distance to the galaxy center, have been accreted further back in time  \citep{Rocha:2012}.

From Figure \ref{fig:velocities} (left), it can be noted that the 
GCs lying farther than 10~kpc are mostly metal-poor,  whereas at smaller distances both kinds of GCs are found. Majority of the GCs have velocities less than $\pm$200~km\,s$^{-1}$.
In the right panel of Figure \ref{fig:velocities}, we show the velocity versus distance along the major axis. 
In this plot it is possible to observe that most of the GCs follow a pattern of disk rotation, i.e. they present a gradient in velocity typical of rotating systems.


\begin{figure}
    \includegraphics[width=0.9\columnwidth]{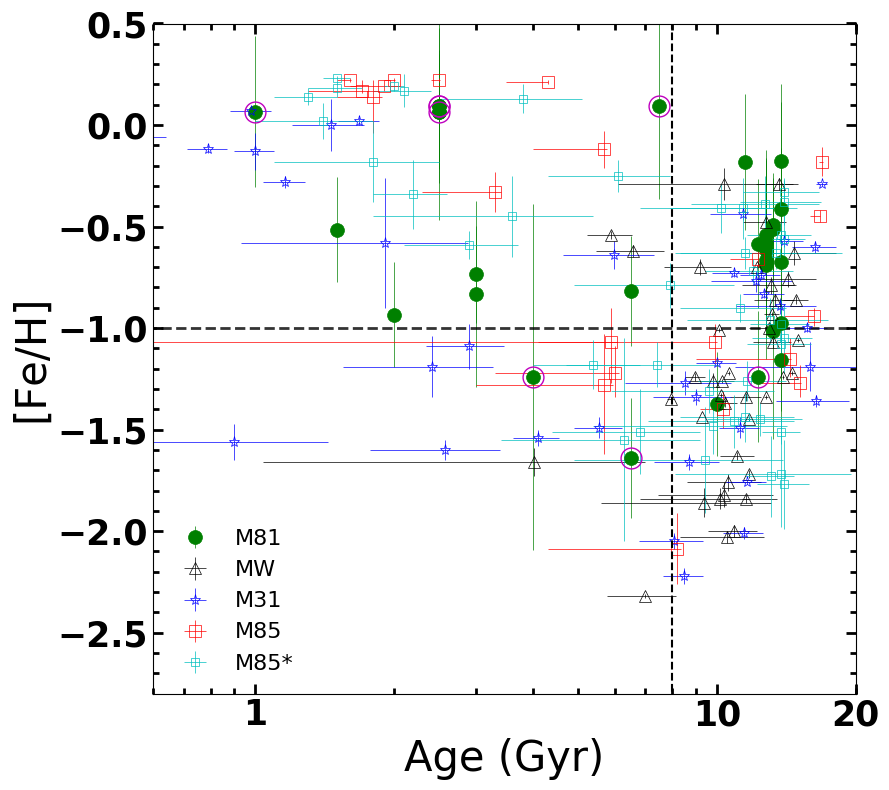}\\
    \includegraphics[width=0.9\columnwidth]{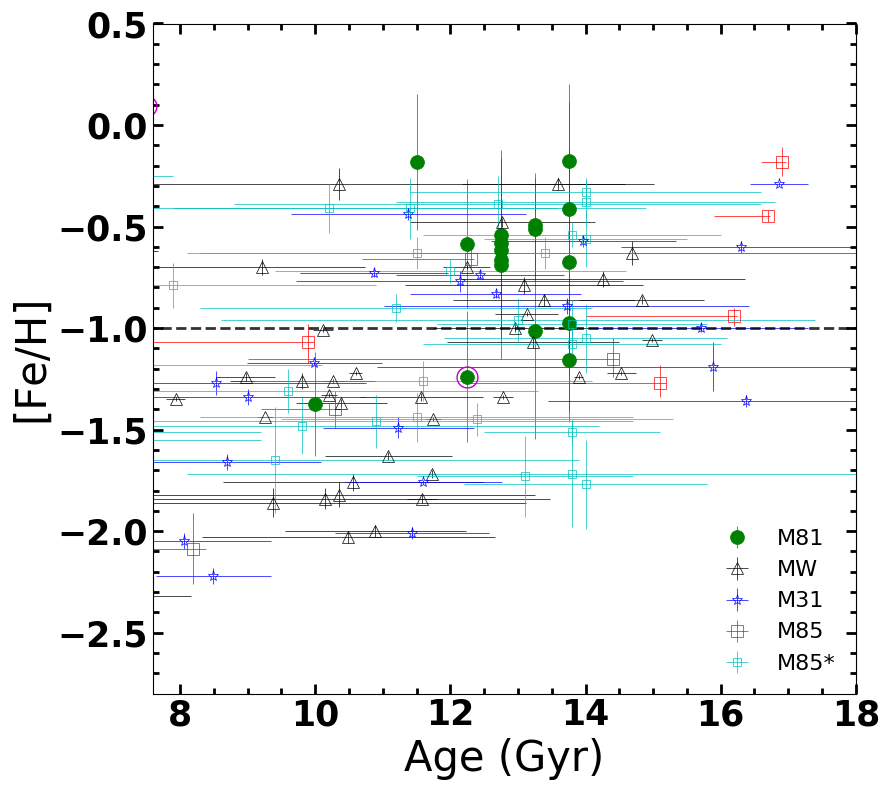}
    \caption{
    Metallicity vs age for different galaxies: M81 ({\it green solid points}), MW ({\it black empty triangles}), M31 ({\it blue empty stars}), M85 ({\it red empty squares}) and M85$^{*}$ ({\it cyan empty squares}). 
    In both panels the {\it horizontal black dashed line} indicates the limit between metal-rich ([Fe/H]$> -1.0$) 
    and metal-poor ([Fe/H]$\leq -1.0$) GCs.  
    In top panel the {\it vertical black dashed line} indicates the limit between intermediate and old age clusters. 
    {\it Top:}  X-axis in logarithmic scale for all the range of ages. The GCs whose metallicity was estimated using the grid method from Figure~\ref{figura:hbeta_mgfe} are circled in magenta.
    {\it Bottom:} X-axis in linear scale with ages from 7.6 to 18 Gyr.
    }
    \label{fig:metalicidad_edad}
\end{figure}

\begin{figure}
    \includegraphics[width=1.0\columnwidth]{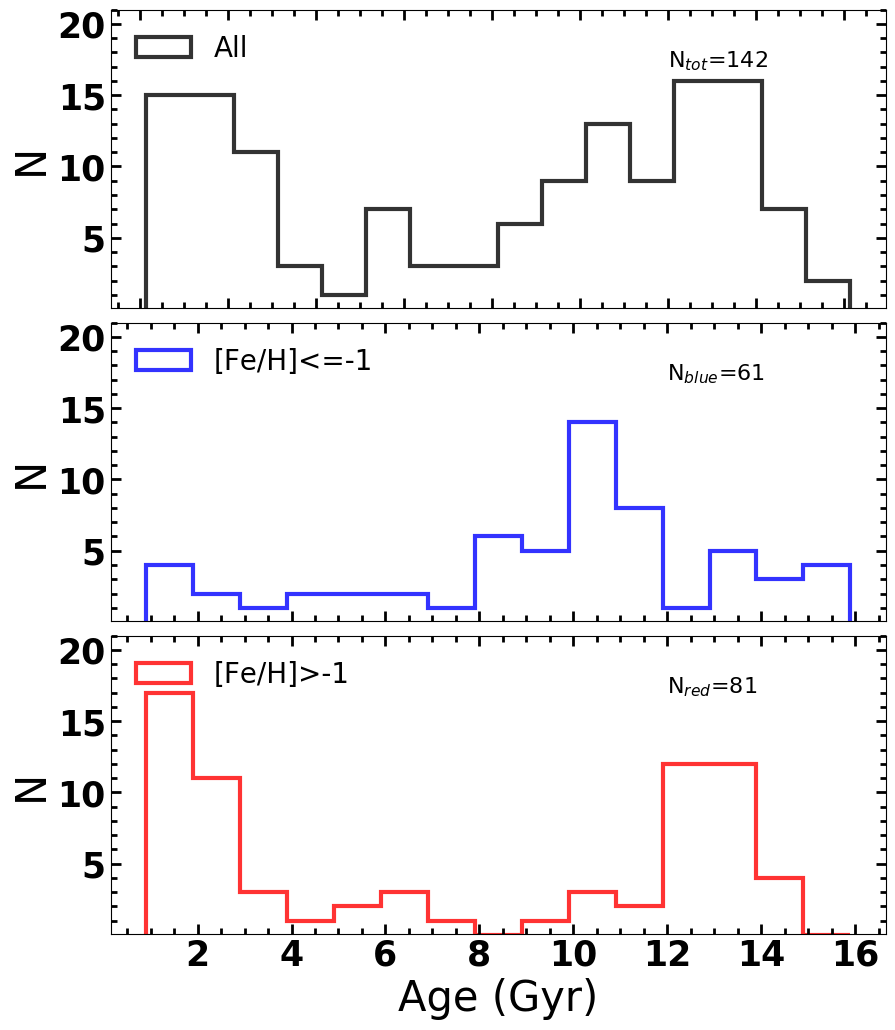}
    \caption{
    {\it Top:} age distribution for  all GCs in M81, MW, M31 and M85 galaxies. 
    {\it Middle:}  age distribution for metal-poor GCs ([Fe/H]<=-1). 
    {\it Bottom:} age distribution for metal-rich GCs ([Fe/H]>-1). 
    In the three panels N is the number of GCs.}
    \label{fig:distribucion_edad}
\end{figure}

\subsection{Cosmological implications from metallicity vs age diagram for spiral and lenticular galaxies}
\label{seccion:metallicity_age}

Most of the age determinations for extragalactic GCs are based on the analysis of colours and Spectral Energy Distributions (SEDs), which have the advantage of studying the properties for the whole population. However, photometrically-based quantities suffer from age, reddening and metallicity degeneracy, which makes the obtained results unreliable. 
On the other hand, the availability in recent years for Multi-Object Spectroscopic (MOS) and/or Integral Field Unit (IFU) facilities at large telescopes covering reasonably large FoVs have enabled the determination of the ages without the effects of degeneracy. For example, the main part of our analysis is based on three MOS pointing observations each covering a FoV of $7\times2$~arcmin$^2$ ($7.5\times6.0$~arcmin$^2$) at the 10.4-m GTC.

We combine our measurements with spectroscopic measurements for other galaxies from the literature and analyse them in a metallicity-age diagram in Figure \ref{fig:metalicidad_edad}. The top panel shows the ages on a logarithmic scale, whereas the bottom panel is zoomed in on a linear scale for ages $>$8~Gyr to show metallicity-age trends for classical GCs. The sources of literature data are: MW (empty black triangles)  and M31 (empty blue stars)  data from \citet{Cezario:2013}, where they used the ULySS package \citep{Koleva:2009} to derive the metallicity and age determination, and M85 GCs (empty red squares) from \citet{Ko:2018}, where they used different methods for estimating age and metallicity, but for consistency, we only show the results that they obtained with the ULySS package. Also we show the M85 GCs (empty cyan squares) from \citet{Escudero:2022}, who also used ULySS package to derive the metallicity and age determination. We refer to the latter sample as M85*.
A horizontal dashed line divides metal-poor (\meta$\leq-$1.0) GCs from the metal-rich (\meta$>-$1.0) GCs.

It can be noticed from the figure that the extragalactic GCs with spectroscopic ages are dominated by the metal-rich population, with  most of the measurements for the metal-poor GCs coming from the Galactic (from a sample of 41 GCs, 19 have an Age$>$~8~Gyr and \meta$\leq-$1.0)
and M85* (from a sample of 46 GCs, 14 have an Age$>$~8~Gyr and \meta$\leq-$1.0) GCs, whose spectroscopic sample covered halo GCs faraway to the galaxy center. This is due to an observational bias originated due to a tendency to maximize the objects per pointing, which invariably leads to select zones of high object density for spectroscopic observations. Such strategies end up sampling the disk GCs which are metal-rich. Differently, from other works, in \citet{Escudero:2022} the outer part of the galaxies were observed, combining part of the disk and possibly halo of the galaxy. Interestingly, some of the GCs selected in that work present halo like characteristics, metal-poor and older than 12 Gyr. We can draw some general remarks from the combined sample.

Metal-poor GCs are concentrated in the age range from 8 to 12~Gyr, with a mean age that is $\sim$3~Gyr younger than that for metal-rich GCs. According to the Hierarchical scenario of galaxy formation \citep[e.g.][]{Cote:1998, Strader:2005, Forbes:2018}, metal-poor GCs are formed in the halos of low-mass galaxies, which are then accreted onto the giant galaxies during subsequent merger processes.
Under this scenario, metal-poor GCs are expected to be the oldest GCs \citep[e.g.][]{Peng:2004, Beasley:2008} 
which is opposite to what is inferred from the figure. The observed contradiction 
is opposite to what 
is expected if the spectroscopic determinations for some unknown reason suffer from age-metallicity degeneracy. On the other hand, the presence of metal-rich old (12--13~Gyr) GCs could be understood under the Cosmological context as it takes less than a 1~Gyr for metal enrichment in GCs. 
To better illustrate these statements, in the Figure~\ref{fig:distribucion_edad} we show the GCs age distributions. In the top panel of Figure~\ref{fig:distribucion_edad} we show the spectroscopic age distribution of  142 GCs, for M81, MW, M31 and M85 galaxies. An age bimodality  can be seen at young and old ages. In the middle panel, the mean age distribution for metal-poor GCs ([Fe/H] $<=-$1), whereas the corresponding  value for metal-rich ([Fe/H] $>-$1) GCs is 12.8~Gyr. The metal-rich GCs also show a peak at relatively a young age of 2~Gyr.

How do we reconcile the absence of metal-poor GCs among the oldest GCs? We believe the answer lies in the observational bias discussed above. The accreted old metal-poor GCs are expected to be in the outer halos of galaxies. In general, these outer-halo GCs are not included in the spectroscopic samples, which explains the absence of old metal-poor GCs in Figure \ref{fig:metalicidad_edad}. 
The work of \citet{Escudero:2022} mitigates this situation by finding several old, metal-poor GCs in the outer part of M85, suggesting that interesting insights on galaxy formation and its connection with its GC system will arise by an extensive study of GCs in galaxy outskirts, as well a systemic compilation of archival data.
The presence of metal-poor, but marginally younger GCs among the spectroscopic samples suggests that not all the metal-poor GCs are formed at the earliest epochs. This can be explained under the hierarchical scenario of galaxy formation, if GCs formation in some of the accreting low-mass galaxies took place after a delay of a few gigayears. Another possibility is that these metal-poor GCs are formed {\it in situ} when gas-rich galaxies merge with a giant galaxy. The fact that these metal-poor GCs are found in spectroscopic samples that cover the inner parts of galaxies favours the later hypothesis. The acceptance of this hypothesis requires a change in the current paradigm that all {\it in situ} formed GCs are metal-rich. Surprisingly there are very few metal-rich GCs of age between 12 and 8~Gyr. As well as, the formation of metal-poor GCs seems to have stopped suddenly at $\sim$8~Gyr.

M81 GCs follow another trend seen in other galaxies, namely the presence of intermediate-age clusters that have photometric and morphological characteristics similar to the classical GCs. Metal-rich objects dominate the sample at these ages. The metal-poor GCs are scarce at ages younger than 8~Gyr. This suggest that if metal-poor GCs in the inner disk are formed {\it in situ} during accretion of gas-rich dwarf galaxies, such process stopped suddenly at $\sim$6~Gyr after the Big Bang. This implies that most of the accreting galaxies have sufficiently enriched their metal content at this epoch. The accretion process and formation of GC-like clusters continued after that, but the newly formed clusters were mostly metal-rich.

\section{Conclusions}

In this work, we presented the results of the analysis of 
new spectroscopic data of 
42 GC candidates in the nearby spiral galaxy M81, obtained with GTC-OSIRIS in the long-slit and MOS modes. 
Spectra for 30 of these objects have a S/N $>$ 10 providing an accurate determination of metallicity, age and extinction.
We used the classical \hb\ vs MgFe diagram to separate clusters that are younger than $\sim$3~Gyr. For objects older than 3~Gyr, we used the iron indices (Fe52, Fe53 and Fe54) to determine  the metallicity. We used the $\chi^2$-technique to determine the best-fit SSP model age of each GC candidate. We find that 17 GC candidates are genuine classical GCs with ages $>$8~Gyr. The remaining 13~candidates (43\%) are intermediate-age clusters. M81 GCs experience a mean $A_V=0.80\pm0.13$~mag
which illustrates the importance of reddening corrections while inferring any physical quantity from optical colours and/or SEDs. We combine the spectroscopically determined age and metallicity of M81 GCs with those obtained using similar techniques for other samples of nearby galaxies to discuss the trends seen in the metallicity-age diagram. 

Most of the metal-rich ([Fe/H] $>-$1) GCs in spectroscopic samples of spiral galaxies are either older than 12 Gyr, or younger than 8 Gyr, with the age range of 8 to 12 Gyr being populated by metal-poor GCs. We postulate that these metal-poor GCs were formed in situ in the disks of galaxies during the accretion of gas-rich dwarf galaxies. Clusters younger than 8~Gyr are systematically metal-rich, suggesting the lack of accreting metal-poor dwarf galaxies at these relatively recent epochs. The absence in spectroscopic samples of old ($>$12~Gyr) metal-poor GCs is most likely due an observational bias of not including outer halo GCs, which are expected to be metal-poor. We recommend spectroscopic studies to include more outer halo GC candidates to address this question.

\section*{Acknowledgements}

We thank an anonymous referee whose inputs were very useful in improving the discussion section.
LLN thanks CONACyT for granting PhD research fellowship that enabled him to carry out the work presented here. We also thank CONACyT for the research grants CB-A1-S-25070 (YDM), and CB-A1-S-22784 (DRG). 
LLN thanks Funda\c{c}\~ao de Amparo \`a Pesquisa do Estado do Rio de Janeiro (FAPERJ) for granting the postdoctoral research fellowship E-40/2021(280692)

This work has made use of data from the European Space Agency (ESA) mission {\it Gaia} (\url{https://www.cosmos.esa.int/gaia}), processed by the {\it Gaia} Data Processing and Analysis Consortium (DPAC, \url{https://www.cosmos.esa.int/web/gaia/dpac/consortium}). Funding for the DPAC has been provided by national institutions, in particular the institutions participating in the {\it Gaia} Multilateral Agreement.

Facility: GTC.

Software: GALAXEV \citep{Bruzual:2003}, GMM \citep{Muratov:2010}, GTCMOS \citep{Mauricio:2016}.

\section*{Data availability}

The paper is based on publicly available archival data. The processed fits files and data tables will be shared on reasonable request to the first author.




\bibliographystyle{mnras}
\bibliography{example} 




\appendix

\newpage

\newpage
\bsp	
\label{lastpage}
\end{document}